\documentclass[sigconf,nonacm,balance=false]{acmart}
\usepackage[most]{tcolorbox}
\usepackage{array}
\usepackage{arydshln}
\usepackage{pifont}
\usepackage{subcaption}
\usepackage{float}
\usepackage{enumitem}
\usepackage[dvipsnames]{xcolor}
\usepackage{xspace}

\newcommand{\Ord}{\mathsf{Ord}}
\newcommand{\HCS}{\mathsf{HCS}}
\newcommand{\Env}{\mathcal{E}}
\newcommand{\Obs}{\mathcal{O}}
\newcommand{\Acts}{\mathsf{Act}}

\newcommand{\View}{\mathsf{View}}

\newcommand{\KL}{\mathrm{D_{KL}}}
\newcommand{\MH}{\texttt{Maude-HCS}}

\begin{document}

\title{\MH: Model Checking the Undetectability-Performance Tradeoffs of Hidden Communication Systems}


\author{Joud Khoury}
\affiliation{%
  \institution{RTX BBN Technologies}
  \city{Cambridge}
  \state{MA}
  \country{USA}}
\email{joud.khoury@rtx.com}

\author{Minyoung Kim}
\affiliation{%
  \institution{SRI International}
  \city{Menlo Park}
  \state{CA}
  \country{USA}}
\email{minyoung.kim@sri.com}

\author{Christophe Merlin}
\affiliation{%
  \institution{RTX BBN Technologies}
  \city{Cambridge}
  \state{MA}
  \country{USA}}
\email{christophe.merlin@rtx.com}

\author{Jos\'{e} Meseguer}
\affiliation{%
  \institution{U. of Illinois at Urbana-Champaign}
  \city{Urbana}
  \state{Illinois}
  \country{USA}}
\email{meseguer@illinois.edu}

\author{Zachary Ratliff}
\affiliation{%
  \institution{Harvard University}
  \city{Cambridge}
  \state{MA}
  \country{USA}}
\email{zacharyratliff@g.harvard.edu}

\author{Carolyn Talcott}
\affiliation{%
  \institution{SRI International}
  \city{Menlo Park}
  \state{CA}
  \country{USA}}
\email{carolyn.talcott@sri.com}


\renewcommand{\shortauthors}{Khoury et al.}

\begin{abstract}
Hidden communication systems (HCS) embed covert messages within ordinary network activity to hide the presence of communication. In practice, the \emph{undetectability} of an HCS is typically evaluated using ad hoc traffic statistics or specific detectors, making security claims tightly coupled to experimental setups and implicit adversarial assumptions. 
In this work, we formalize undetectability as the statistical indistinguishability of observable execution traces under two deployments: a baseline system without hidden communication and an HCS deployment carrying covert traffic. Undetectability is expressed as a bound on a quantitative measure of distance between the trace distributions induced by these two executions.

We develop \MH, an executable modeling and analysis framework that provides a principled and executable foundation for reasoning about undetectability–performance tradeoffs in complex HCS designs.
\MH\ allows designers to specify protocol behavior, adversary observables, and environmental assumptions, and to generate Monte Carlo samples from the induced trace distributions. We demonstrate that \MH\ can be used to audit claims of undetectability by estimating the true and false positive rates of a statistical test and converting these estimates into lower bounds on undetectability measures such as KL divergence. This enables systematic evaluation of detectability and its tradeoffs with performance under explicitly stated modeling assumptions.

Finally, we evaluate \MH\ on proof-of-concept tunneling-based HCS instantiations and validate model predictions against measurements from a physical testbed. For passive adversaries observing timing and traffic statistics, we quantify how undetectability and performance vary with protocol configuration, background traffic, and network loss, and demonstrate strong semantic alignment between model-based guarantees and empirical results. 
\end{abstract}


\maketitle

\section{Introduction}\label{sec:intro}
Hidden communication systems (HCS) embed covert messages into ordinary network activity, with the goal of concealing not only the content of communication but its very existence. 
Such systems are critical for both the internet freedom and the national security communities~\cite{dingledine_tor:_2004, DARPA_RACE, DARPA_PWND2}.
For example, they enable journalists (and warfighters) to covertly coordinate and circumvent online censorship in authoritarian and/or adversarial environments. 
The use of these systems carries significant risk, where {\em detection of ``fact of communication''} can result in loss of life~\cite{Dorfman2018}.

A broad range of hidden communication systems have been proposed and/or deployed. 
The techniques used by these systems may be broadly classified into protocol tunneling (e.g.,~\cite{sun2023telepath, wails2022learning, rosen2021balboa, barradas2020poking, barradas2017deltashaper, mcpherson2016covertcast}), mimicry (e.g.,~\cite{mohajeri2012skypemorph, dyer2015marionette, dyer2013protocol}), and/or obfuscation (e.g., ~\cite{Yawning_obfs4, winter2013scramblesuit, wails2023proteus}); reflection/refraction (e.g.,~\cite{vandersloot2020running, nasr2017waterfall, karlin2011decoy, ellard2015rebound}); cloud fronting and proxying (e.g.,~\cite{zolfaghari2016practical, fifield2015blocking}); and steganography (e.g.~\cite{meadows2022predicting, TRISTvinod}). 
An important property common to these systems is how they balance {\em performance} with risk of {\em detection}.
Different techniques occupy different areas in the undetectability-performance solution space.
Both performance (e.g., the time it takes to send a message of a certain size) and risk of detection by the adversary are of paramount importance to the HCS user, whose needs are generally driven by mission requirements (e.g., a fixed window of time to deliver a message of a certain size at an acceptable level of risk).
%
%

Although hidden communication systems have been extensively studied in the literature, no general and scalable framework or tool exists today for formally modeling and reasoning about their performance and undetectability properties.
Validating undetectability and performance properties of HCS designs today remains largely informal, relying on simulations and testbed experimentation of the HCS implementations.
While this approach has been very valuable to date, it is rigid, error-prone, and takes significant time and effort.
In practice, this lack of formal models of HCSs (including their implicit and explicit assumptions) and of a principled and modular approach for reasoning about their properties makes it hard to effectively explore the design and parameter space, to evolve the design, or to compare different designs.

On the other hand, state-of-the-art techniques have significantly advanced the theory and practice of formalizing distributed network system designs and reasoning about their security~\cite{barbosa2021sok, blanchet2012security, cortier2011survey, santiago_formal_2014, cheval_indistinguishability_2023, meier2013tamarin} and performance properties~\cite{liu_formal_2023, DBLP:journals/pacmpl/LiuMOZB22}. 
These formal approaches, however, have not been applied to the HCS domain, and they remain very tedious in practice requiring highly skilled experts. 
In particular, many design choices must be made to choose suitable formalisms, abstractions, analysis frameworks and properties, and scaling techniques.


We present \MH, {\em the first generalized toolchain for formally specifying and reasoning about hidden communication systems at real-world scales}.\footnote{\href{https://github.com/raytheonbbn/maude-hcs}{https://github.com/raytheonbbn/maude-hcs}} 
The toolchain, comprised of a language and analysis toolkit, enables network designers to model and explore alternative HCS designs rapidly and effectively, and validates privacy-performance guarantees needed to trust the design --- a necessary step for HCS users to ultimately trust the system, especially when operating in high threat environments. 

\MH\ brings state-of-the-art formal tools and frameworks to the HCS domain.
The basis of the toolchain is Maude’s expressive, extensible, and executable language based on rewriting logic~\cite{unified-tcs}. Rewriting logic, with its real-time and probabilistic rewrite theories~\cite{agha_pmaude_2006, journ-rtm, clavel_all_2007, clavel_maude_2024}, is ideally suited for formalizing distributed object-oriented systems such as HCSs. The formalism is timed and probabilistic, thus able to represent pertinent behaviors and properties of HCSs and adversaries hunting for them (e.g., stochastic message sizes, arrivals, delays, computation time, timers and timeouts, traffic and behavior models, and randomized algorithms).
The formalism is also general and faithful, so it can express a wide range of hiding techniques and applications without compromising fidelity. The formalism is executable with high performance,
enabling in-situ, fast, and automated formal analysis of HCSs.

To enable scalable and rigorous analysis, the \MH\ toolkit supports verification of precise quantitative privacy and performance properties, expressed in the expressive QuaTeX quantitative and probabilistic temporal logic~\cite{agha_pmaude_2006, gul-survey} using Statistical Model Checking (SMC) tools in Maude and QMaude~\cite{DBLP:conf/fm/RubioMPV23, fadoss_umaudemc}.\footnote{The toolchain additionally supports qualitative security and privacy properties expressed in modal and temporal logic by standard model checking.} 
SMC reasons about precise quantitative properties while remaining scalable by estimating statistical quantities up to a configurable confidence level using Monte Carlo simulations~\cite{agha_pmaude_2006, gul-survey}. 
%


This work makes the following key contributions.
\begin{itemize}
\item {\bf Formalizing undetectability over observable trace distributions}.
We formalize {\em undetectability} as a quantitative indistinguishability condition between the distributions of observable execution traces induced by the ordinary system (without hidden communications) and the HCS (with hidden communication). Undetectability is expressed as a bound on a divergence measure between these trace distributions. 
This formulation makes explicit the dependence of undetectability on \emph{observable} trace semantics, and provides a precise language for stating quantitative undetectability claims.

\item {\bf Modular semantically-aligned models and scalable analysis.}
\MH\ integrates state of the art tools and techniques~\cite{DBLP:journals/pacmpl/LiuMOZB22, timedPTran, meseguer_taming_2014, DBLP:conf/fm/RubioMPV23} 
to reduce the modeling and analysis burden on HCS designers. 
We build on the state-of-the-art modular approach of Liu et al.~\cite{DBLP:journals/pacmpl/LiuMOZB22} for formally specifying concurrent distributed systems with time and probabilities as generalized actor theories~\cite{DBLP:conf/birthday/000126}.
Designers specify nondetermnistic models, which are automatically transformed using sound theory transformations into probabilistic versions that are semantically aligned.
We leverage a recent important extension~\cite{timedPTran} that allows specification of timed models as input to the transformation, to allow faithfully expressing inherent semantics in HCSs such as timers and timeouts.
To further encourage generality and modularity of HCS specifications, we systematize different classes of HCSs as formal patterns~\cite{galan_protocol_2023, meseguer_taming_2014}.
The tunneling pattern for instance embeds hidden data in legitimate cover protocol sources.\footnote{The specification of the cover protocol is independent of the tunneling protocol; the latter wraps the former with additional functionality without modifying it.}
The approach significantly reduces the burden on the designer for implementing HCS models so they can focus on modeling the novel aspects of their HCS design, while ensuring semantic consistency of the overall model, and thus increases trust in the analysis results.

\item {\bf Methodology for auditing undetectability claims.}
We introduce a methodology for auditing divergence-based undetectability claims for concrete HCS designs using SMC. 
Given a protocol, a specified adversary observation model, and a claimed privacy parameter, \MH\ enables designers to sample observable traces under both ordinary and HCS executions and to compute adversary-relevant statistical quantities under each model. 
These measurements allow practitioners to assess whether the claimed divergence bound is supported under the specified assumptions and to explore how undetectability varies as system parameters (e.g., adversary detection logic or application performance) are modified.
We demonstrate the predictive power of our approach by comparing model-based guarantees to actual experimental results on a controlled testbed, and showing strong alignment between the two.
\item \textbf{Scalable auditing across parameterized deployment regimes.} We demonstrate that Maude-HCS enables systematic auditing of undetectability–performance tradeoffs by sweeping deployment and adversary parameters. Across multiple scenarios, we vary application rate, network loss, background traffic, and detector thresholds, and compute measures of undetectability for each operating point. This shows that the tool supports structured exploration of large design spaces rather than single-point evaluation.
\end{itemize}
The paper is organized as follows. 
Section~\ref{sec:background} introduces relevant background material. 
Section~\ref{sec:hcs-privacy} introduces our formal definition of undetectability over observable trace distributions. 
Section~\ref{sec:instrumentation} describes our pipeline and methodology for modeling HCS systems and auditing their properties.
Section~\ref{sec:evaluation} presents a thorough evaluation of the effectiveness of the approach for auditing performance-privacy claims, and highlights the predictive power of the approach relative to experimental results.
Finally, we conclude in Section~\ref{sec:conclusion}.

\section{Background and Related Work}
\label{sec:background}

This section provides the relevant background on the probabilistic modeling and analysis framework used in this paper, and reviews related work.

To reason about distributional properties of distributed systems we rely on the PMaude \cite{agha_pmaude_2006} probabilistic modeling framework and its associated statistical model checking capabilities. PMaude builds on Maude’s rewriting logic semantics for concurrent and distributed systems, extending it with time and probability.

\vspace{1.5ex}

\noindent {\bf Rewriting Logic, Maude and Real-Time Maude}.  Rewriting logic~\cite{unified-tcs} is a computational
logic well suited to specify concurrent systems.  A rewrite theory 
$\mathcal{R}$ specifies:
(i) a system's \emph{distributed states}
as an \emph{algebraic data type}, and (ii) its \emph{local concurrent transitions}
by rewrite rules\footnote{More generally, rewrite rules can be \emph{conditional}, i.e., of the
form $u \rightarrow u' \; \mathit{if} \; \mathit{cond}$, so that the rule does not fire
unless the condition or guard $\mathit{cond}$ is satisfied.  Henceforth
all rules will be silently assumed to be possibly conditional even if $\mathit{cond}$ is not mentioned.}
$u \rightarrow u'$, where $u$ is a pattern describing certain state fragments, and $u'$
is the pattern describing the resulting local state after the transition.
For example, the distributed state of a
 \emph{generalized actor rewrite theory} \cite{DBLP:conf/birthday/000126}  $\mathcal{R}$ is
a multiset of objects and messages, and its rewrite rules specify its asynchronous communication events such as,
for example, a server's response to a client's message, formalized by
a rule of the form $(x \lhd y:m)\; \langle x \mid atts\rangle \rightarrow \langle x \mid atts'\rangle \; (y\lhd x:m')$,
where $\langle x \mid atts\rangle$ is the current state $atts$ of sever $x$
when receiving a message $m$ from client $y$, which then changes its state to $atts'$
and sends back a message $m'$ to $y$. More generally, the rules in a generalized actor rewrite theory
are either: (i) \emph{message-triggered rules} of the form\footnote{For simplicity, actors are here 
described as objects  $\langle o \mid atts\rangle$, where $o$ is the actor's name or identifier,
and its state $\mathit{atts}$ is a record of attribute-value pairs.  But this is an \emph{untyped} view of actors, which
can more generally be of the form $\langle o: C \mid atts\rangle$, where $C$ is the actor's \emph{class},
and actor classes are related by class inheritance hierarchies \cite{ooconc}.}
$(x \lhd y:m)\; \langle x \mid atts\rangle \rightarrow \langle x \mid atts'\rangle \; \mathit{msgs}$, where $\mathit{msgs}$
is a (possibly empty) set of messages sent by $x$,
which capture the Actor model \cite{agha_actors_1986,DBLP:conf/birthday/000126}, or (ii)
\emph{object-triggered rules} of the form 
$\langle x \mid atts\rangle \rightarrow \langle x \mid atts'\rangle \; \mathit{msgs}$,
where an actor on its own changes its state and sends messages.  For example, the rule's left-hand side
$\langle x \mid atts\rangle$ may be a pattern describing the expiration of a timer in $x$, which
may be reset and may cause some messages to be sent.
Maude \cite{maude-book} is a high-performance implementation of rewriting logic
as a declarative programming language.  Concurrent systems with real-time features such as local clocks, timers and timestamps
are modeled and analyzed as \emph{real-time rewrite theories} in the Real-Time Maude language extension \cite{journ-rtm}.

\vspace{1.5ex}

\noindent {\bf Probabilistic Rewrite Theories, PMaude, and Statistical Model Checking}.
Generalized actor systems are intrinsically nondeterministic, both because of their asynchronous behavior
and because of unpredictable  network delays.  \emph{Probabilistic rewrite theories}
\cite{agha_pmaude_2006,DBLP:journals/pacmpl/LiuMOZB22} make it possible to \emph{quantify} the non-determinism of such systems as follows.  
Message delays in a given network
infrastructure can be modeled by a continuous probability distribution $\pi$ that agrees with experimental measurements.
Then, our above client-server rule becomes the \emph{probabilistic rewrite rule}
$(x \lhd y:m)\; \langle x \mid atts\rangle \rightarrow \langle x \mid atts'\rangle \; \mathit{delay}((y\lhd x:m'),d)
\; \mathit{with} \; \mathit{probability}\; d := \pi$.  That is, the nondeterministic delay $d$ in the transmission of message
$(y\lhd x:m')$ becomes now probabilistically quantified by the distribution $\pi$ from which $d$ is sampled.
$\mathit{delay}((y\lhd x:m'),d)$ describes a message \emph{en route} to $y$, which ``arrives'' to $y$, i.e., it becomes
$(y\lhd x:m')$,
when the system's global clock advances $d$ time units.
PMaude \cite{agha_pmaude_2006} extends Real-Time Maude to support \emph{probabilistic (and real-time)
generalized actor rewrite theories} of this kind.
Since these theories quantify the system's real-time behavior, they make it possible to measure various system performance
properties, such as latency, throughput, and so on.  Furthermore, such quantitative properties can be systematically estimated
by submitting the P-Maude generalized actor rewrite theory $\mathcal{R}$ to \emph{statistical model checking} (SMC) \cite{gul-survey}.

However, for SMC to be possible some requirements must be met.  First, probabilistic
rewrite rules are not directly executable.  For example, the choice $d := \pi$ in the above 
client-server rule is not executable: it must be \emph{simulated} by sampling $\pi$.  In \cite{DBLP:journals/pacmpl/LiuMOZB22}
$\mathcal{R}$ is made executable by the $\mathit{Sim}$ transformation $\mathit{Sim}: \mathcal{R} \mapsto \mathit{Sim}(\mathcal{R})$, which
makes SMC possible on $\mathcal{R}$ by performing \emph{Monte Carlo simulations}
of $\mathit{Sim}(\mathcal{R})$.  However, for most SMC tools, $\mathcal{R}$ must be a \emph{purely probabilistic}
model.  This means that in any Monte-Carlo simulation of $\mathcal{R}$ the probability of reaching a
state where two different transitions can be nondeterministically chosen is $0$.  This is called the
\emph{absence of nondeterminism} (AND) property \cite{DBLP:journals/pacmpl/LiuMOZB22}.

The AND property problem can be solved by solving a bigger problem, namely, automating the passage from real-time actor
rewrite theories to PMaude rewrite theories.  For example, automating the passage from a client-server system
to its PMaude counterpart.  This can be done for any real-time actor rewrite theory $\mathcal{R}$
provided a specification $\Pi$ assigning a probability distribution $\pi_{r}$
to each rule $r$ in $\mathcal{R}$ is given.  This is the so-called $P$ \emph{transformation}
$P: \mathcal{R} \mapsto \mathcal{R}_{\Pi}$ in \cite{DBLP:journals/pacmpl/LiuMOZB22}, where $\mathcal{R}_{\Pi}$ is a PMaude actor rewrite theory.
As shown in \cite{DBLP:journals/pacmpl/LiuMOZB22}, under mild conditions on the input theory
 $\mathcal{R}$ and on the continuous distributions in $\Pi$, $\mathcal{R}_{\Pi}$ satisfies the AND
 property and is therefore amenable to SMC analysis. 
 There is, however, a remaining problem: the input theories
 in \cite{DBLP:journals/pacmpl/LiuMOZB22} are \emph{untimed}, i.e., they do not have any real-time features
 like clocks, timers, timeouts, etc.  This serious limitation\footnote{The limitation is serious because
 many theories $\mathcal{R}$ in PMaude, including those with real-time features, cannot be expressed
 as theories of the form $\mathcal{R}=\mathcal{R}_{0_{\Pi}}$ for $\mathcal{R}_{0}$
 an untimed generalized actor rewrite theory.}
 has been recently overcome in \cite{timedPTran} by: (i) 
 generalizing $P$ to the \emph{timed} $P$ transformation $P: \mathcal{R} \mapsto \mathcal{R}_{\Pi}$, where
 now $\mathcal{R}$ can be a real-time generalized actor rewrite theory, and (ii) proving that,
 under mild conditions on 
 $\mathcal{R}$ and $\Pi$, $\mathcal{R}_{\Pi}$ satisfies the AND property.
 
 There are still two more issues in the SMC verification of PMaude theories.
 First, we need a probabilistic temporal logic to express quantitative properties.  QuaTex \cite{agha_pmaude_2006}
 is an expressive \emph{quantitative} probabilistic logic.  In standard temporal logic state predicates are 
 Boolean-valued.  In QuaTex state properties are \emph{real-valued}, so that when model checking 
 a QuaTex formula $\varphi$
 in a PMaude theory, the result is the estimated value of the property specified by $\varphi$, e.g., latency or throughput, etc.
   Second, it may not be possible to evaluate sophisticated QuaTex properties in the models $\mathcal{R}_{\Pi}$
 or $\mathit{Sim}(\mathcal{R}_{\Pi})$ because the relevant information has not been ``logged'' in the states. For example, 
 estimating the average delay of all queries in a distributed transaction system requires access to such logs.
 The user can modify $\mathcal{R}_{\Pi}$
 or $\mathit{Sim}(\mathcal{R}_{\Pi})$ to log relevant events; but this is tedious and error-prone.  The $M$ transformation
in \cite{DBLP:journals/pacmpl/LiuMOZB22},
 $M : \mathit{Sim}(\mathcal{R}_{\Pi}) \mapsto M(\mathit{Sim}(\mathcal{R}_{\Pi}))$ automates this last step
 in a correct-by-construction way.  The $P$, $\mathit{Sim}$ and $M$ transformations automate
 the entire SMC process of model checking a QuaTex qualitative property on a PMaude theory.
 QMaude \cite{DBLP:conf/fm/RubioMPV23} is at present the most efficient SMC tool to verify QuaTex properties 
 on PMaude theories.  An example illustrating all the ideas in this section, as well as the use of the
 PMaude pipeline for analyzing the undetectability of HCS systems, is given in Appendix \ref{sec:rtt-section}.

In this work we use the PMaude pipeline to define executable probabilistic semantics for both ordinary and HCS deployments, and to generate samples of observable traces under each model. We leverage statistical model checking to estimate adversary-relevant quantities, such as the true positive rate and false positive rate of a specified detection test, together with confidence intervals. These estimates are then used to derive lower bounds on divergence measures, enabling us to audit claimed undetectability guarantees as described in Section~\ref{subsec:undetectability-auditing}.

\vspace{1.5ex}

\noindent {\bf Related Work} Recent work has argued that hidden communication systems should be evaluated using formal cryptographic methodologies rather than ad hoc empirical testing. In particular, Pereira et al.~\cite{pereira2025position} propose modeling undetectability as simulation-based indistinguishability between real HCS traces and ideal benign traces and advocate machine-checked proofs of such claims. Our work is complementary. We formalize undetectability as an information-theoretic property of the trace distributions induced by a specified deployment model and develop Maude-HCS as a framework for auditing such claims under concrete system and adversary assumptions. Rather than proving undetectability universally, we instantiate executable probabilistic models and estimate divergence over parameterized observation windows, enabling quantitative evaluation of detectability and its tradeoffs with performance in realistic deployments.

State-of-the-art frameworks for quantitative analysis of distributed systems are typically based on statistical model checking (SMC)~\cite{gul-survey} or probabilistic model checking (PMC)~\cite{baier_model_2018}. 
Maude-HCS adopts an SMC-based approach over executable rewriting-logic models, trading the exactness of PMC for improved scalability while retaining precise quantitative reasoning through probabilistic temporal logic and Monte Carlo estimation with confidence guarantees. 
In contrast to automata-based tools such as UPPAAL-SMC~\cite{david_uppaal_2015}, PRISM~\cite{kwiatkowska_prism_2011}, SBIP~\cite{mediouni_bip_2018}, and the Modest toolset~\cite{hartmanns_modest_2014, budde_jani_2017}, which rely on finite-state models and often require ad hoc encodings to represent distributed protocols with dynamic actors, message queues, and unbounded data structures, Maude-HCS operates directly on expressive actor-style rewrite theories. This enables faithful modeling of concurrent networked systems and supports both qualitative and quantitative analysis within a single semantically aligned executable specification, avoiding the modeling inconsistencies that can arise when separate abstractions are used for different forms of analysis.
\section{Formalizing Undetectability of Hidden Communication Systems}
\label{sec:hcs-privacy}
At a high level, an HCS seeks to embed messages into ordinary network activity in such a way that an external observer cannot reliably determine whether hidden communication is taking place. Conceptually, there are two worlds of execution
\begin{itemize}
    \item an \emph{ordinary} world $\Ord$ in which the baseline protocol executes without hidden communication
    \item a \emph{hidden communication} world $\HCS$ in which the same protocol carries covert messages
\end{itemize}

Undetectability is defined in terms of the interactions between an observer and each of these worlds.

\paragraph{Adversary model.}
We let $\Obs$ denote the alphabet of protocol observables such as packet sizes, timings, and directions. 
We let $\Acts$ denote the set of actions available to the adversary such as dropping packets, delaying messages, or influencing protocol behavior. 
We let $\Env$ denote the set of environment states such as network queues, CPU load, and background traffic conditions.

An adversary $\mathcal{A}$ is an interactive possibly randomized algorithm that, given a transcript prefix over the protocol's execution, outputs the next action in $\Acts$.

\paragraph{Interactive execution.}
Execution is modeled as an interactive protocol between a world $X \in \{\Ord,\HCS\}$, an environment state $E_t \in \Env$, and an adversary $\mathcal{A}$. At time $t=0$, the environment is initialized to some state $E_0 \in \Env$. 

At each time step $t$, the adversary selects an action $A_t \in \Acts$, and the world evolves according to its transition semantics and produces an observable $O_t \in \Obs$ together with an updated environment state $E_{t+1}$.

A transcript up to time $t$ is the sequence
\begin{align*}
\pi_{0:t}
=
((E_0, A_0, O_0), (E_1, A_1, O_1), \ldots, (E_t, A_t, O_t))
\end{align*}

\paragraph{Adversary view and induced trace distributions.}
Fix a world $X \in \{\Ord,\HCS\}$, an adversary $\mathcal{A}$, and an initial state $E_0 \in \Env$. 
For each such choice, the interaction between $X$ and $\mathcal{A}$ over horizon $t$ defines a probabilistic experiment where we let $\View^{X}_{\mathcal{A},E_0,t}$
denote the random variable corresponding to the adversary’s \emph{view} of the interaction, defined as the projection of the full transcript onto the observable interaction:
\begin{align*}
\View^{X}_{\mathcal{A},E_0,t}
=
\{(A_i,O_i)\}_{i=0}^{t}
\subset \pi_{0:t}
\end{align*}

\par\smallskip

\paragraph{Undetectability.}
Hidden communication is undetectable if no adversary interacting with either world can reliably distinguish which world generated its observed transcript.

\begin{definition}[$(\mathcal{M}, d)$-Undetectability]
\label{def:undetectability}
Fix an observation horizon $t$. Let $\mathcal{M}$ be a divergence or distance measure on pairs of probability distributions over transcripts of length $t$, and let $d$ be a parameter in the range of $\mathcal{M}$. A deployment satisfies $(\mathcal{M}, d)$-HCS undetectability up to horizon $t$ if $\forall \mathcal{A},\ \forall E_0 \in \Env$,
\begin{align*}
\mathcal{M}\big(
\View^{\HCS}_{\mathcal{A},E_0,t},
\View^{\Ord}_{\mathcal{A},E_0,t}
\big)
\le d
\end{align*}
\end{definition}

This definition admits multiple operational interpretations depending on the choice of $\mathcal{M}$.\footnote{We do not require $\mathcal{M}$ to be symmetric. Many divergences with meaningful operational interpretations are inherently asymmetric. The ordering $\mathcal{M}\big(\View^{\HCS}_{\mathcal{A},E_0,t}, \View^{\Ord}_{\mathcal{A},E_0,t}\big)$ reflects the detection-oriented setting in which the adversary seeks evidence that an observed transcript was generated by the HCS world rather than the ordinary baseline.} For example, when $\mathcal{M}$ is KL divergence $\KL(P\|Q)=\mathbb{E}_{x\sim P}\!\left[\log \frac{P(x)}{Q(x)}\right]$, $(\mathcal{M},d)$-Undetectability bounds the adversary’s expected log-likelihood ratio, that is, its expected information gain from observing a length-$t$ trace. For divergence measures that admit hypothesis-testing interpretations, classical inequalities relate divergence to testing performance. In particular, when $\KL\big(\View^{\HCS}_{\mathcal{A},E_0,t} \,\|\, \View^{\Ord}_{\mathcal{A},E_0,t}\big)$ is small, no detector can simultaneously achieve a low false positive rate and a substantially higher true positive rate. Other probability distance measures, such as total variation distance or max divergence, admit analogous operational guarantees.

\subsection{Auditing Undetectability via Monte Carlo Trace Sampling}
\label{subsec:undetectability-auditing}

Definition~\ref{def:undetectability} characterizes undetectability in terms of a divergence $\mathcal{M}(Q_t,P_t) = \mathcal{M}(\View^{\HCS}_{\mathcal{A},E_0,t}, \View^{\Ord}_{\mathcal{A},E_0,t})$ between the trace distributions induced by the ordinary and HCS worlds, quantified over all adversary strategies and initial environmental conditions. We show how \MH\ can be used to \emph{audit} claims of undetectability by providing a high-probability lower bound on $\mathcal{M}(Q_t,P_t)$ from Monte Carlo samples. Such an audit is necessarily carried out for a specific instantiated model, adversary, execution horizon $t$, and distribution over initial environmental conditions, as determined by the executable semantics of the system under consideration. In this work we focus on KL divergence as the undetectability measure, though the same methodology applies to other divergence notions such as max divergence and total variation distance. A certified lower bound on $\KL(Q_t\|P_t)$ can be used directly to audit undetectability claims: if a deployment claims $(\KL,d)$-HCS undetectability, meaning that $\KL(Q_t\|P_t) \le d$, then any certified lower bound exceeding $d$ falsifies the claim under the specified modeling assumptions, whereas a lower bound below $d$ does not refute it.

We emphasize that this methodology provides only a lower bound on $\KL(Q_t\|P_t)$. Passing such an audit does not establish an upper bound on the true divergence, nor does it prove that the claimed privacy parameter holds universally. Rather, it provides evidence consistent with the claim under the sampled executions, adversary model, and environmental assumptions encoded in the executable semantics. \\

\noindent\textbf{Auditing Framework}. Although undetectability is defined in terms of divergence between trace distributions, directly estimating these distributions is infeasible in many practical settings because observable traces are inherently high dimensional. For example, a passive adversary may observe the full sequence of inter packet arrival times over a window of length $t$, producing a vector whose dimension depends on the number of packets transmitted. Directly estimating the full distributions $P_t$ and $Q_t$ over such spaces is statistically infeasible, since discretizing a $d$-dimensional observable with $B$ bins per coordinate yields a joint domain of size $B^d$. As a result, the number of Monte Carlo samples required to reliably estimate divergence becomes prohibitively large.

In this regime we instead audit detectability through an explicit adversarial test. Let $\varphi \colon \mathcal{O}_t \to \{0,1\}$ be any detector that attempts to distinguish HCS traces from ordinary traces. Its false positive and true positive rates are
\begin{align*}
\mathrm{FPR}_\varphi &= \mathbb{E}_{o \sim P_t}[\varphi(o)] \\
\mathrm{TPR}_\varphi &= \mathbb{E}_{o \sim Q_t}[\varphi(o)]
\end{align*}
Let $Y = \varphi(O_t)$ denote the detector's one bit output. Under $P_t$, the random variable $Y$ is distributed as a Bernoulli random variable $\mathrm{Bern}(\mathrm{FPR}_\varphi)$, and under $Q_t$ it is distributed as $\mathrm{Bern}(\mathrm{TPR}_\varphi)$. By the data processing inequality for KL divergence,
\begin{align*}
\KL(Q_t \| P_t)
\ge
\KL\big(\mathrm{Bern}(\mathrm{TPR}_\varphi) \,\|\, \mathrm{Bern}(\mathrm{FPR}_\varphi)\big)
\end{align*}
Equivalently,
\begin{align*}
\KL(Q_t \| P_t)
\ge
\mathrm{TPR}_\varphi \log \frac{\mathrm{TPR}_\varphi}{\mathrm{FPR}_\varphi}
+
(1-\mathrm{TPR}_\varphi) \log \frac{1-\mathrm{TPR}_\varphi}{1-\mathrm{FPR}_\varphi}
\end{align*}

Estimating $\mathrm{TPR}_\varphi$ and $\mathrm{FPR}_\varphi$ from $n$ Monte Carlo samples in each world yields confidence intervals
\[
\mathrm{TPR}_\varphi \in [T_L,T_U],
\qquad
\mathrm{FPR}_\varphi \in [F_L,F_U]
\]
that hold jointly with probability at least $1-\delta$. A conservative high-probability lower bound is obtained by minimizing the Bernoulli KL divergence over all parameters consistent with these intervals. Since $\KL(\mathrm{Bern}(q)\|\mathrm{Bern}(p))\ge 0$ with equality iff $p=q$, this minimum is attained by choosing $p$ and $q$ as close as possible. In particular, with probability at least $1-\delta$, we have:
\begin{align*}
\KL(Q_t \| P_t) &\ge 
\min_{q\in[T_L,T_U],\,p\in[F_L,F_U]} \KL(\mathrm{Bern}(q)\|\mathrm{Bern}(p)) \\
&=
\begin{cases} 
0 & \text{if } [T_L,T_U]\cap[F_L,F_U]\neq\emptyset,\\[2pt]
\KL(\mathrm{Bern}(T_L)\|\mathrm{Bern}(F_U)) & \text{if } T_L > F_U,\\[2pt]
\KL(\mathrm{Bern}(T_U)\|\mathrm{Bern}(F_L)) & \text{if } T_U < F_L
\end{cases}
\end{align*}
We report this minimized value as our certified lower bound. \\

\noindent\textbf{Operational interpretation.}
The KL divergence $\mathrm{KL}(Q_t \| P_t)$ equals the adversary's expected log-likelihood ratio in favor of the HCS world after observing a length-$t$ trace. If $\mathrm{KL}(Q_t \| P_t) = \varepsilon$, then the adversary's posterior odds increase by a factor of $e^\varepsilon$ in expectation. For example, if the adversary assigns prior probability $0.01$ to HCS activity, then the prior odds are $0.01/0.99 \approx 0.0101$. If $\varepsilon = 1$, the expected posterior odds become $0.0101 \cdot e \approx 0.0275$, corresponding to posterior probability approximately $0.027$. Thus, KL divergence directly quantifies the expected multiplicative shift in posterior odds induced by a single length-$t$ observation window.

\section{Modeling and Analysis Workflow}
\label{sec:instrumentation}
\begin{figure*}[t]
    \centering
    \includegraphics[width=\linewidth]{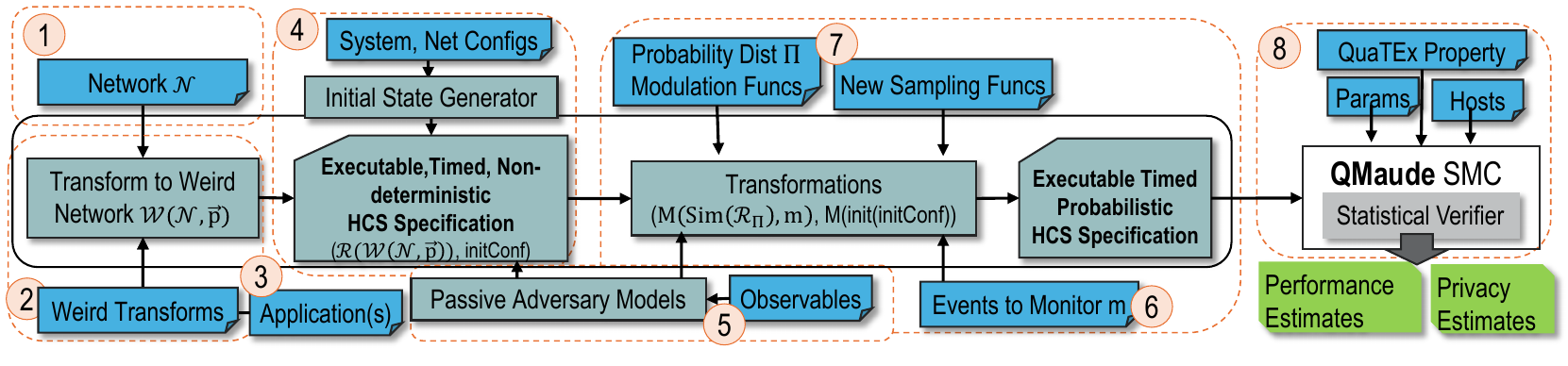}
    \caption{\MH\ reduces the burden on the designer while ensuring semantic consistency. Designers build nondeterministic models of the underlying network \ding{172}, of the weird transforms \ding{173}, and of the application \ding{174} as generalized actor theories. They also provide the actor and system configs \ding{175} which are used to generate the initial HCS configuration. Designers then specify the adversary observables \ding{176} (and extend the adversary detection logic as needed), and the performance relevant events to monitor \ding{177}. \MH\ generates executable, timed, and probabilistic specification used for statistical analysis. For analysis, designers specify the quantitative performance and privacy properties of interest and the SMC parameters $\alpha$, $\delta$ (and hosts for distributed SMC) \ding{179}. These are used to produce performance and privacy guarantees.}
    \label{fig:workflow}
\end{figure*}
\MH\ enables designers to model their HCS design, and to evaluate performance and privacy guarantees about their model.
Designers can ask, for example, {\em what is the goodput of a hidden communication system design for a desired probability of detection?}

The modeling and analysis workflow is shown in Figure~\ref{fig:workflow}.
The model of the underlying network and protocols without hidden communication, $\mathcal{N}$, is directly formalized in Maude as a generalized actor rewrite theory that can be analyzed as
an independent  entity.
A formal pattern~\cite{meseguer_taming_2014} transformation $\mathcal{N}\mapsto \mathcal{W}\left(\mathcal{N},\vec{p}\right)$ takes the underlying network system $\mathcal{N}$ and transforms it to a new weird network system.\footnote{We refer to the resulting hidden communcation system as a weird network.}
The parameters $\vec{p}$ typically include a set of one or more transformations that embed/extract hidden information. 
The network system $\mathcal{N}$ is preserved as a subsystem of the weird network $\mathcal{W}\left(\mathcal{N},\vec{p}\right)$.
The $\mathcal{N}\mapsto \mathcal{W}\left(\mathcal{N},\vec{p}\right)$
transformation wraps each selected actor object in $\mathcal{N}$ inside a corresponding meta-actor, which hides its communications with other objects in $\mathcal{N}$ {\em without changing the original object’s behaviors}.  
Using the $M$ transformation~\cite{DBLP:journals/pacmpl/LiuMOZB22},
a monitor object records selected events during execution that are relevant for computing by SMC
performance measures specified as QuaTex form.
Similarly, we implement an adversary object that records selected observables during execution that are relevant for detectability 
analysis based on a model of the adversary's capabilities.
Monitored events and adversary observables are programmed using patterns (e.g., filtering the messages the passive adversary can observe).
Appendix~\ref{sec:rtt-section} illustrates this modeling approach using a simple HCS example.

As discussed in Section~\ref{sec:background}, an HCS model begins with a non-deterministic timed rewrite theory $R$, which is transformed into a probabilistic timed rewrite theory $M(\mathit{Sim}(\mathcal{R}_{\Pi}))$, in which sources of uncertainty, such as message delays, are modeled using explicit probability distributions, and distributions are simulated via sampling. 
As a result, $\mathit{Sim}(\mathcal{R}_{\Pi})$ defines a well-defined probability measure over execution traces. 
There is no residual nondeterminism, and each execution is fully determined by the sampled random values. \\


\noindent\textbf{Ordinary and HCS worlds.}
Our modular modeling approach allows us to easily distinguish between two executable systems.
The ordinary system is represented by $M(\mathit{Sim}(\mathcal{R}(\mathcal{N})_{\Pi}))$, and does not contain the weird transforms or the HCS application actors (this is the system without hiding).
On the other hand, the HCS system is represented by $M(\mathit{Sim}(\mathcal{R}(\mathcal{W}(\mathcal{N},\vec{p}))_{\Pi}))$ and includes the weird network behaviors and the HCS application.
These two rewrite theories share the same baseline protocol structure and differ only in whether hidden communication is embedded. 
Both systems are instrumented with identical monitors, so that they expose observables in a common domain $\mathcal{O}$.

Executing each system samples a probability distribution over $\mathcal{O}_t$. We denote these distributions by $P_t$ for the ordinary system and $Q_t$ for the HCS system. These distributions are induced by the probabilistic semantics of the corresponding rewrite theories, including any randomized initialization of system and environment state specified by the model.
\par\smallskip
\noindent\textbf{Undetectability}. 
At a high level, \MH\ serves two purposes. First, it defines a precise notion of what a passive adversary can observe with its observational 
capabilities by explicitly instrumenting executions to record observable traces. Second, it enables sampling from the induced distributions over those traces, which can then be compared using statistical techniques. Much of the underlying infrastructure already exists in PMaude and QMaude. Our contribution in this paper
is in developing the theoretical foundations of undetectability analysis for HCS systems, and in extending the Maude-HCS
infrastructure to support privacy analysis, where the object of interest is inherently relational and distributional.
Rather than checking temporal properties of a single system execution, we are interested in comparing the \emph{distributional behavior} of two related systems: an ordinary deployment and an HCS deployment. \MH\ provides a workflow for instrumenting executable probabilistic models so that such comparisons can be carried out in a principled and repeatable manner.

Once instrumented, each model is analyzed using statistical model checking. Rather than exporting raw traces and performing offline estimation, we specify adversary-relevant quantities—such as the probability that a given detector alarms within horizon $t$—as quantitative QuaTEx properties. QMaude then estimates these probabilities via Monte Carlo simulation with user-specified confidence parameters. In particular, for a fixed detector $\varphi$, we directly obtain high-confidence estimates of $\mathrm{TPR}_\varphi$ and $\mathrm{FPR}_\varphi$ under the HCS and ordinary models, respectively. These SMC-derived estimates are then used to compute the certified lower bound on $\KL(Q_t \| P_t)$ described in Section~3.1. In this way, divergence-based undetectability claims are audited through statistically verified detector behavior within each executable model.

\par\smallskip
\noindent\textbf{Performance} 
In the context of a HCS, quantitative performance measures such as latency, goodput, and throughput can be precisely defined
as QuaTex formulas.
For example, if HCS user Alice uses the HCS to transfer a file of size $B$ bytes to user Bob and it takes $T$ seconds, the goodput is $B/T$ bytes/second. 
These quantitative measures can be precisely specified using QuaTex in \MH.
\MH\ SMC provides a statistical guarantee of the form {\em the expected value of the transfer time is within a range $[\hat{v}-\delta/2, \hat{v}+\delta/2]$ with confidence $(1-\alpha)$}, given desired confidence parameters $\alpha$ and $\delta$~\cite{agha_pmaude_2006}.
Throughout the evaluation, we always set $\alpha=0.05$ (95\% confidence).
The designer can thus characterize the performance of the design, by assessing the effect of the different design parameters on performance.
For example, {\em how does link loss (or MTU size) affect file transfer time?}
This type of performance characterization can yield insights for improving the design, and informing the deployment.

\section{Evaluation}\label{sec:evaluation}

We conduct a thorough evaluation of our approach by instantiating a hidden communications system with 
reasonable complexity, quantifying undetectability and performance properties across a wide range of scenarios and adversary capabilities, and assessing the predictive power of the models by validating the model guarantees against a testbed implementation of the same system.

We structure our evaluation around three sets of experiments. \\ 

\noindent {\bf Privacy-Performance Trade-off Analysis}: Given an adversary detection capability,  we conduct a privacy-performance trade-off analysis across a range of scenarios. 
To do so, we increase application performance and audit KL divergence between ordinary and HCS distributions across a range of observables. \\
        
\noindent {\bf Sensitivity Analysis}: We vary the (1) initial configuration by instantiating different scenarios (e.g., by varying link loss, and background traffic), (2) adversary classifier thresholds, and (3) application parameters (e.g., number of files Alice is sending). We observe the effect on both performance and undetectability. \\

\noindent {\bf Semantic Alignment}: We demonstrate how our model-based results align closely with experimental results from the testbed implementation, attesting to the predictive power of the \MH\ models and guarantees.  

\subsection{HCS System and Experimental Setup}\label{sec:hcs_system_description}
\begin{figure}[t!] 
    \centering 
	\includegraphics[width=0.9\linewidth]{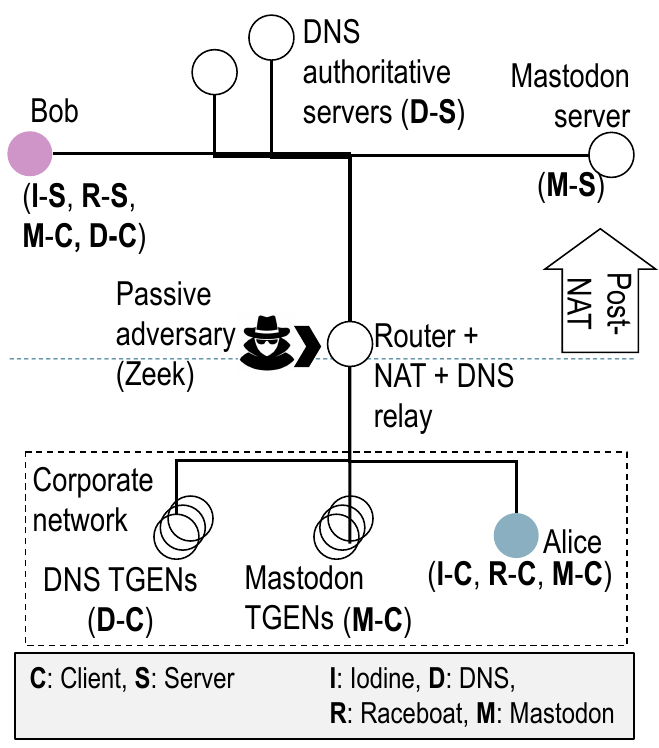}
    \caption{The HCS system and network modeled and instantiated on the testbed, and used for evaluation.}
    \label{fig:testbed_network}
\end{figure}

We model and instantiate a tunneling-based two-channel HCS, comprising of a control channel, and a data channel.
The control channel is realized using Iodine~\cite{iodine} tunneling over the Domain Name System (DNS) protocol~\cite{RFC1034}.
The data channel is realized using Destini~\cite{tstrace_destini, TRISTvinod} tunneling over the Mastodon social network~\cite{Rochko_Mastodon}, the latter serving as a whiteboard. 
Figure~\ref{fig:testbed_network} illustrates the system and the network.
HCS user Alice, located within the corporate network, covertly sends files to Bob located outside the corporate network.

\subsubsection{Testbed Experimental Setup}\label{sec:passive_adversary_description}
The control channel is operated as a QUIC connection over Iodine~\cite{iodine}, a DNS tunneling protocol used to bypass firewalls by encapsulating traffic into DNS queries. We fitted Alice with an Iodine client, and Bob with an Iodine server.
The data channel of the HCS is operated as Destini-over-Mastodon: covert traffic is segmented into byte chunks, which Destini, a steganographic protocol, encodes into cover images. These images are then handled by Raceboat~\cite{tstrace_raceboat} to upload onto a Mastodon server as media attachments in \emph{toots}, which contain hashtags that the server side of the HCS can use to retrieve and then decode covert data. Raceboat acts as a Pluggable Transport that handles encoded and benign image uploads and downloads.

The HCS operates over an underlying network that would look familiar to network operators: Alice and synthetic traffic generation sources (TGENs) operate on a corporate network behind a router that translates network addresses (see dotted box in Figure~\ref{fig:testbed_network}). This is colloquially referred to as NAT, and pre- and post-NAT activity applies to corporate versus all other traffic. Finally, DNS infrastructure is further emulated with public and domain resolvers.

DNS and Mastodon traffic generators mimic real deployments where multiple users utilize common resources, like a Mastodon server or DNS resolvers, alongside the HCS. Background synthetic traffic ends up mixing with HCS traffic and complicates detection. \\

\noindent {\bf Passive Adversary}
The (passive) adversary is located post-NAT as a monitor and can observe all traffic ingressing and egressing the corporate network (from traffic generators and Alice). 
The adversary is instantiated on the testbed using the Zeek monitoring tool~\cite{Paxson1999Bro, ZeekProject}, with a range of detectors.
The detectors were instantiated with thresholds based on baseline data from which covert traffic is absent.
For each detector, when the threshold is exceeded an alarm is recorded in Zeek.
The adversary has three detectors monitoring post-NAT corporate traffic, two are cumulative (C) and one is a moving average (MA).
\begin{itemize}
    \item DNS query count ("C2"): the total number of observed DNS queries is compared to a baseline threshold. 
    \item HTTPS request count ("C8"): the total number of observed HTTPS requests is compared to a baseline threshold. 
    \item DNS query rate ("MA1"): the moving average DNS query rate is compared to a baseline threshold. The query rate is computed over a 60s moving window, starting with baseline queries; when the average rate exceeds $k*R$, where $k$ is a multiplier and $R$ is a well-chosen threshold, for $n$ consecutive bins, the alarm triggers. 
\end{itemize}

\subsubsection{Note on Testbed Implementation}\label{subsec:semantic_alignment:validation_testbed_description}
The testbed matching Figure~\ref{fig:testbed_network} was developed by an independent test and evaluation team implementing all the described functionality, and deployed as a Dockerized network. 
The testbed also recreates the dynamic conditions of a real network, which may experience delays and loss. The testbed instantiated Linux Traffic Control to emulate different link delays (constant, normally distributed around a mean, etc.) and loss (per-packet).
Importantly, the software deployed in the testbed implementation is buggy (e.g., Iodine~\cite{iodine} and Raceboat~\cite{tstrace_raceboat} implementations). 
As we explain in section~\ref{subsec:semantic_alignment}, we identified a range of differences between our model results and the implementations that we attributed to implementation bugs

\subsubsection{Corresponding Actor Models} We developed a formal faithful model of the system, including all the actors for both channels and for the application, and for the underlying network, as well as their configurations.
The Iodine-DNS channel uses DNS as the underlying network; we started with the formal DNS model of~\cite{liu_formal_2023}, and added the Iodine weird network including the Iodine client and server supporting bi-directional communications, as well as the DNS traffic generators.
The Destini-Mastodon channel uses the Mastodon infrastructure as the underlying network where we implement mastodon client and server functionality for posting and fetching toots and images, as well as mastodon background traffic generators.
The weird network was implemented using the raceboat client and server with Destini module covertly encoding application data into images uploaded to mastodon. 

For the HCS application, we implemented Alice as the sending application which encapsulates an iodine client, and a raceboat client, and mastodon client used by raceboat.
Bob is the receiving application, which encapsulates the iodine server and the raceboat server, and a mastodon client used by raceboat.
In terms of infrastructure, we implemented a router that separates the corporate from the public network, and the initial state generator that takes experimental configurations and converts them into an initial configuration for analysis.
Several of the actors actions are driven by {\em markov processes}, including the DNS and mastodon traffic generators, raceboat client and server mastodon clients.
The markov processes have two simple states for posting and fetching media toots. The post or fetch action leads to a wait state during which the underlying Mastodon client undertakes all API-level actions to upload or download media images (encoded or not) and status toots.
Waits in the wait state are normally distributed with a mean and standard deviation specified by the configuration. \\

\noindent {\bf Adversary and Undetectability}. 
We implemented the passive adversary actor including the observables (as patterns over messages).
We implement the three detectors, ``C2'', ``C8'' and ``MA1'', as mixed properties in QuaTEX over the adversary configuration (which captures the trace observables), and quantify their TPR and FPR of each detector by measuring probability of alarm relative to the HCS (TPR) or to the ordinary system (FPR).
In practice, the adversary's detector capability can be abstracted as a binary classifier that takes high dimensional trace observables as input and produces a binary detection decision (and in the case of detection, additional metadata such as the time $t' \leq t$ at which detection occurs). Any such classifier has a TPR and FPR from which we compute a lower bound on KL divergence (Section~\ref{subsec:undetectability-auditing}). 
While in this evaluation we experiment with the three detectors, any such classifier can be audited.

\subsubsection{Quantitative Properties}\label{sec:quanti-properties}
 We measure the following HCS application performance properties, expressed as QuaTEx properties.
 \begin{itemize}
    \item \textbf{Latency:} The delay to deliver the covert files from the moment Alice starts sending the first one, to the moment Alice receives receipt acknowledgment from Bob that the last file was delivered. Note that Alice will not begin to send the next file until it receives an acknowledgment from Bob on the control channel that the previous file was delivered.
    \item \textbf{Goodput:} The total file bytes successfully delivered divided by the latency.
\end{itemize}
The Monitor object in our models keeps track of relevant events required for computing the performance properties. 
We additionally measure the following privacy-related properties, from the perspective of the adversary.
 \begin{itemize}    
    \item \textbf{Operation (Op) duration:} The total experiment time elapsed before the adversary detector alarms. Each detector (C2, C8, MA1) may produce a different op duration value. The op duration is less or equal than than latency when there is a detection (and discarded otherwise). Op duration is naturally expressed in QuaTEx as a {\em mixed property} (example in Appendix \ref{sec:quatex-example}).
    \item \textbf{True Positive Rate (TPR) and False Positive Rate (FPR):} For any input configuration including the adversary detector and its parameters, and resulting trace of length $t$, we measure the probability of alarm on both the ordinary and the HCS systems. These become the FPR and TPR measures, respectively.\footnote{For any sampled trace, if the alarm fires, we measure a 1 and a 0 otherwise.} The TPR and FPR estimates are used to compute the lower bound on KL divergence (section~\ref{sec:hcs-privacy}).
\end{itemize}

\subsubsection{Scenarios}\label{subsec:scenarios_description}
We explore a range of scenarios intended to stress-test and highlight various elements of our model. All scenarios test covert transmissions of files from Alice to Bob. 
Scenarios are defined in groups of three, where all parameters remain constant for the most part, but background traffic is increasingly voluminous.
The first three scenarios have no network loss---they represent somewhat of an idealized network where modest link delays are the only stress.
The following scenarios have various levels of loss between Alice, Bob (which does not communicate directly with Alice), and the Mastodon server and public DNS resolver. Scenarios 4, 5, and 6 have very high loss (5\% packet loss) on the client and server sides of the HCS. Scenarios 7, 8, and 9 have high loss (2.5\%) on each side. Table~\ref{table:scenarios} provide a reference for all noteworthy characteristics of each scenario. In that table, the last column is the number of DNS and Mastodon synthetic traffic generators.
\begin{table}[]
    \centering
    \begin{tabular}{>{\raggedleft\arraybackslash}m{0.95cm} | >{\raggedleft\arraybackslash}m{0.75cm} | >{\raggedleft\arraybackslash}m{0.75cm} | >{\raggedleft\arraybackslash}m{1.0cm} | >{\raggedleft\arraybackslash}m{1.0cm} | >{\raggedleft\arraybackslash}m{1.25cm} }
        Scenario \# & Loss Alice & Loss Bob & Num files & Total Bytes & Num backgrnd generators \\
        \hline
        *1 & 0.0\% & 0.0\% & 10 & 6,820B & 16 \\
        2 & 0.0\% & 0.0\% & 10 & 57,509B & 32 \\
        3 & 0.0\% & 0.0\% & 10 & 57,123B & 48 \\
        \hdashline
        4 & 5.0\% & 5.0\% & 5 & 8,622B & 16 \\
        5 & 5.0\% & 5.0\% & 5 & 20,758B & 32 \\
        6 & 5.0\% & 5.0\% & 5 & 27,486B & 48 \\
        \hdashline
        *7 & 2.5\% & 2.5\% & 5 & 7,463B & 16 \\
        8 & 2.5\% & 2.5\% & 7 & 34,301B & 32 \\
        9 & 2.5\% & 2.5\% & 7 & 39,030B & 48 \\
    \end{tabular}
    \caption{Summary of the test scenarios used for the evaluation audits. *1 and 7 are focus scenarios. Loss Alice is the loss on the link between the router and the mastodon server / public DNS resolver (similarly Loss Bob is loss on the link between Bob and the mastodon server / public DNS resolver.}
    \label{table:scenarios}
\end{table}

\subsection{Privacy-Performance Trade-off Analysis}\label{tradeoff}
\begin{figure*}[t!]
    \centering
    \includegraphics[width=0.33\linewidth]{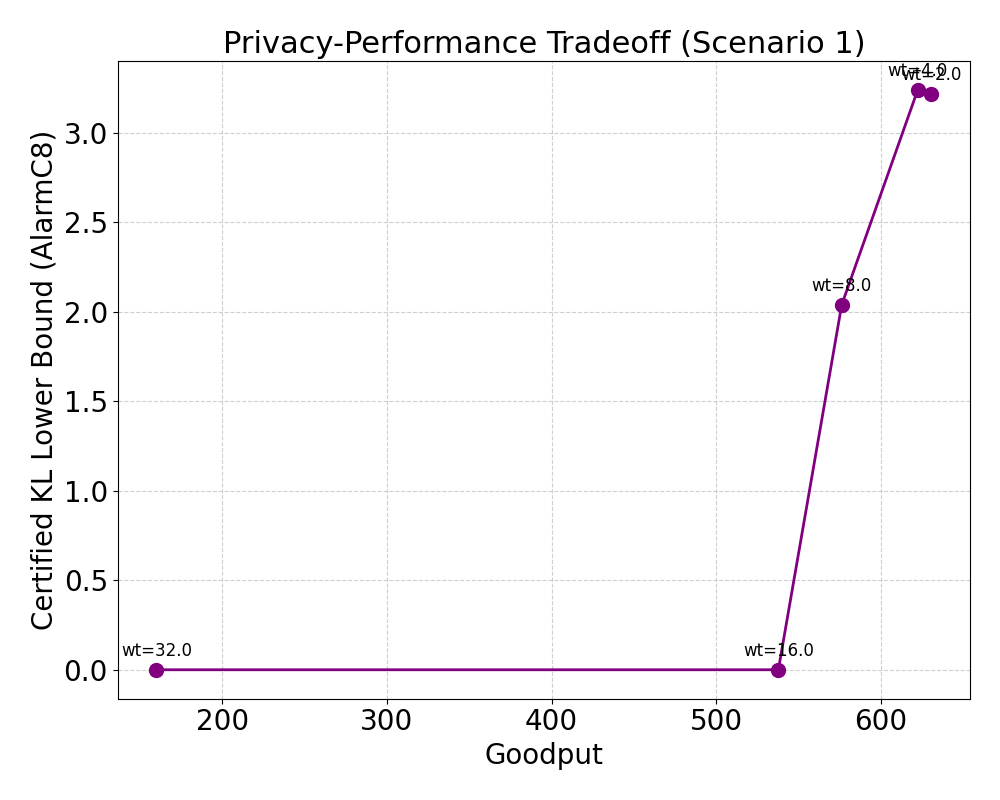}
    \hfill    
    \includegraphics[width=0.33\linewidth]{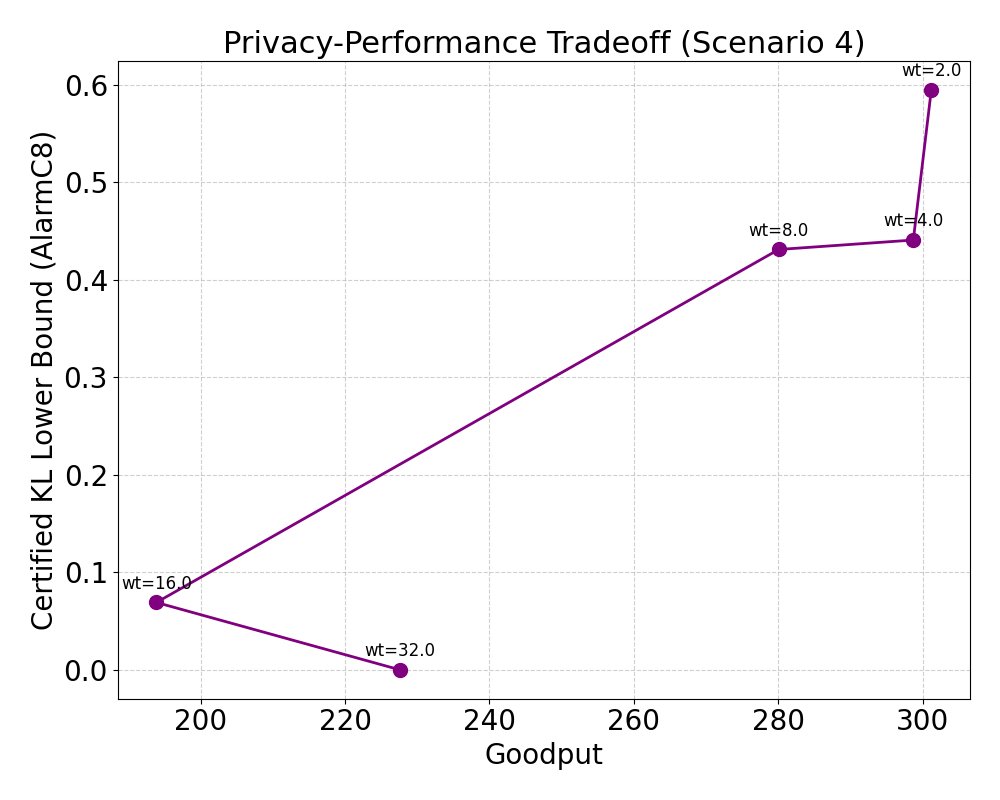}
    \hfill    
    \includegraphics[width=0.33\linewidth]{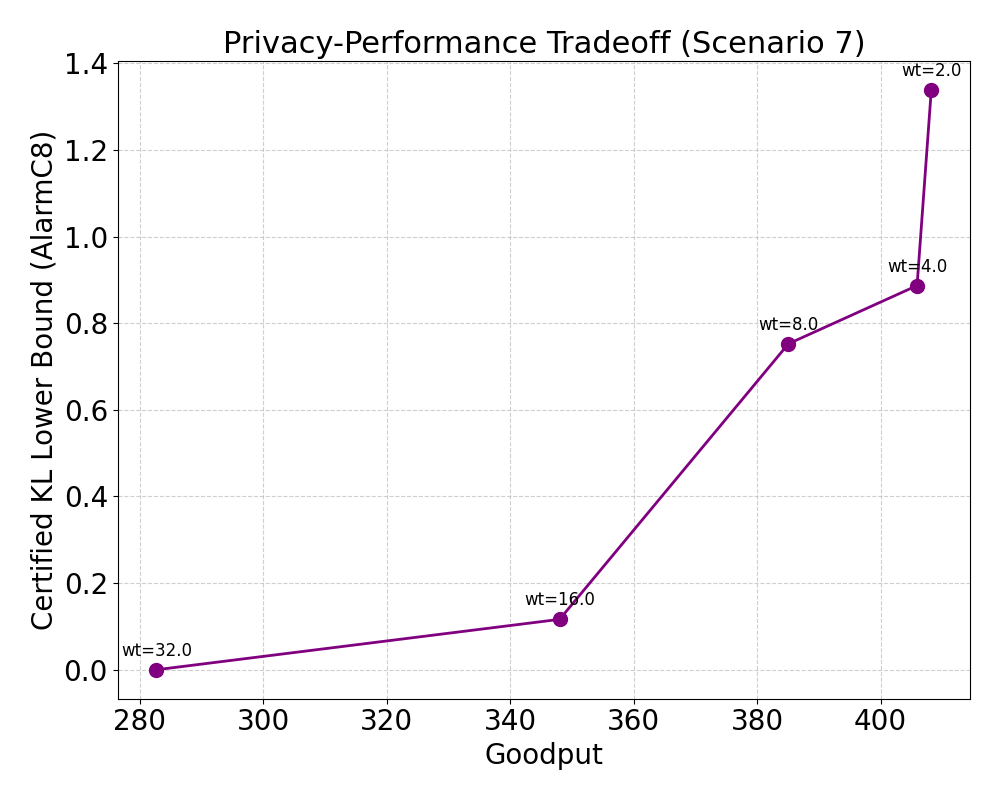}
    \caption{Privacy–performance tradeoff across Scenarios 1, 4, and 7. Each subplot corresponds to a different network condition (see Table 1). For each scenario, Alice’s mean wait time between posts in the Markov model is varied while all other parameters are fixed. Each point shows the achieved goodput (bytes/sec) and the corresponding certified lower bound on $\mathrm{KL}(\View^{\mathrm{HCS}}{t}|\View^{\mathrm{Ord}}{t})$, derived from SMC estimates of detector TPR and FPR. Across scenarios, decreasing the wait time generally increases goodput and generally increases statistical distinguishability between HCS and ordinary executions, though the magnitude of the KL bound depends on network loss and background traffic.}
    \label{fig:tradeoffs-s7}
\end{figure*}
\begin{figure*}[t!]
    \centering    
    \includegraphics[width=0.33\linewidth]{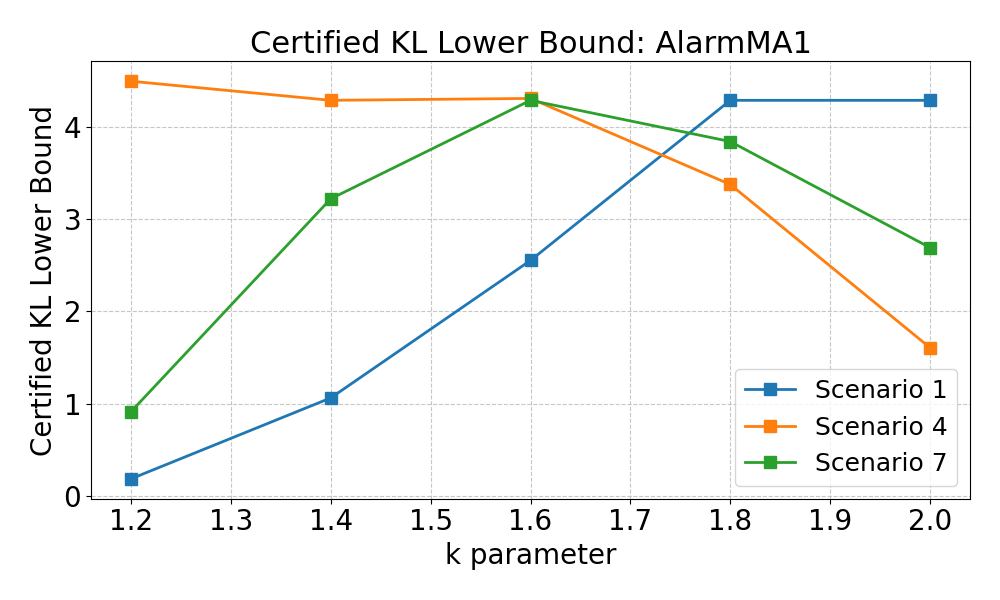}
    \hfill    
    \includegraphics[width=0.33\linewidth]{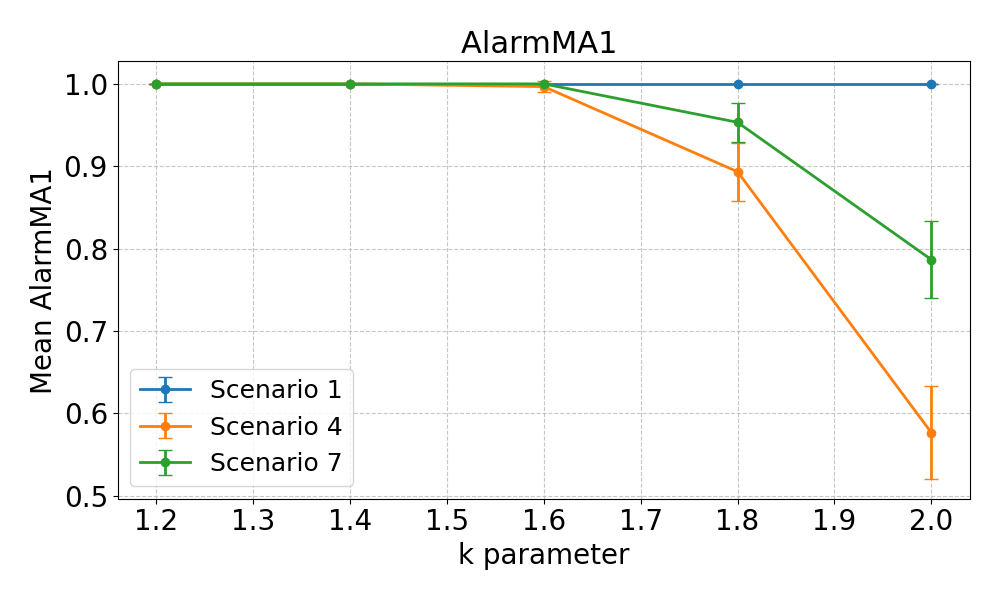}    
    \hfill
    \includegraphics[width=0.33\linewidth]{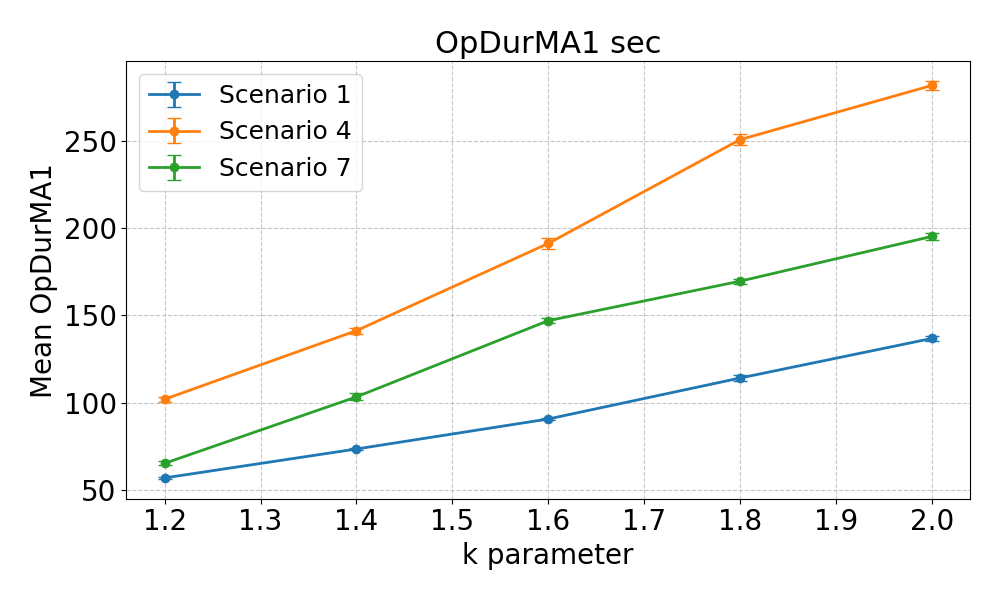}
    \caption{We vary the moving average detector threshold multiplier $k$ (section~\ref{sec:quanti-properties}) and observe the effect on KL divergence bounds (left), on probability of alarm for the moving average detector MA1 (middle), and on operating duration (right). The bar around the alarm datapoints is the radius.}
    \label{fig:tkl-vs-threshold}
\end{figure*}
\begin{figure*}[t!]
    \centering    
    \includegraphics[width=0.33\linewidth]{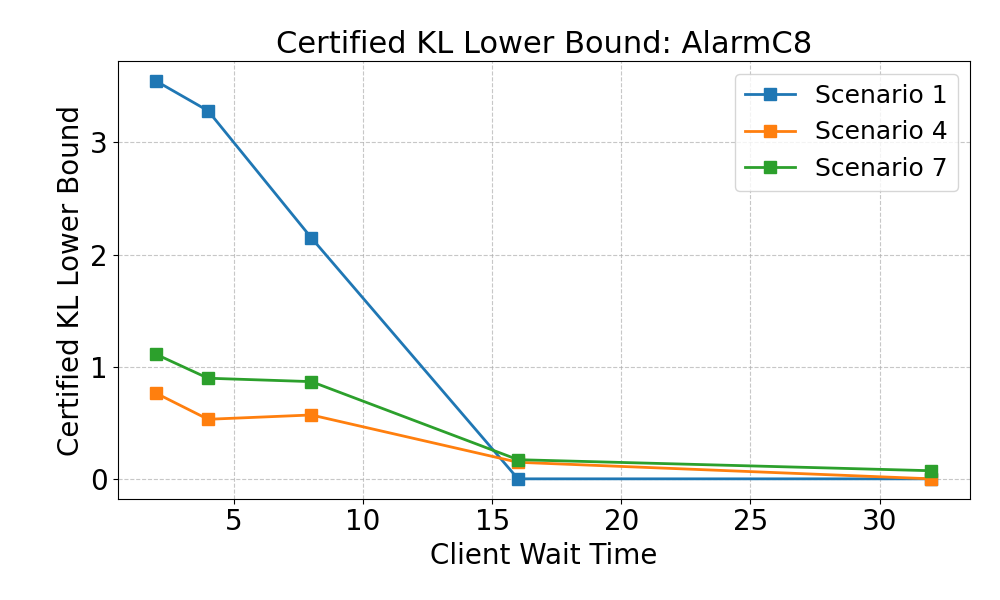}
    \hfill    
    \includegraphics[width=0.33\linewidth]{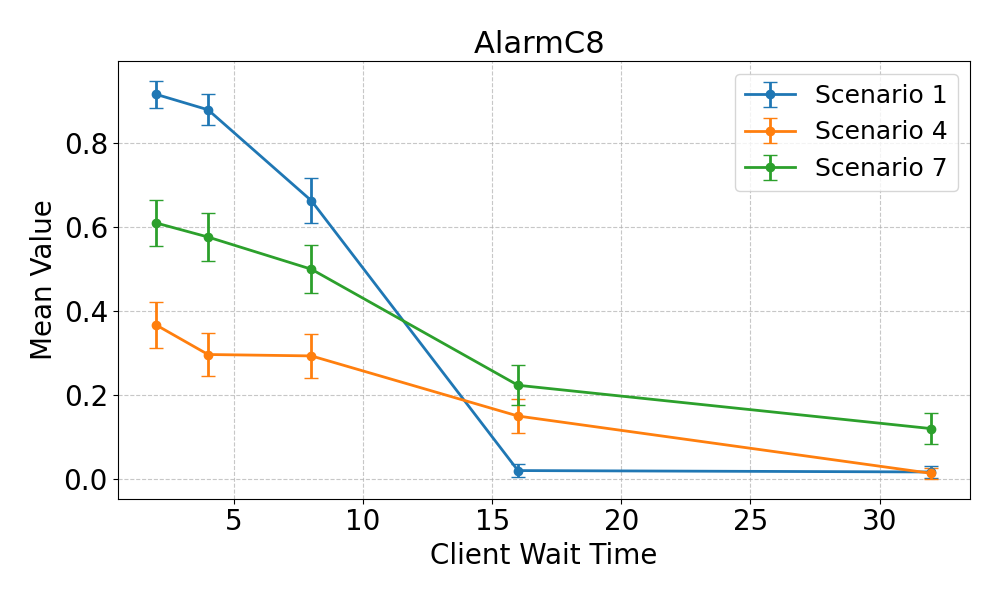}    
    \hfill
    \includegraphics[width=0.33\linewidth]{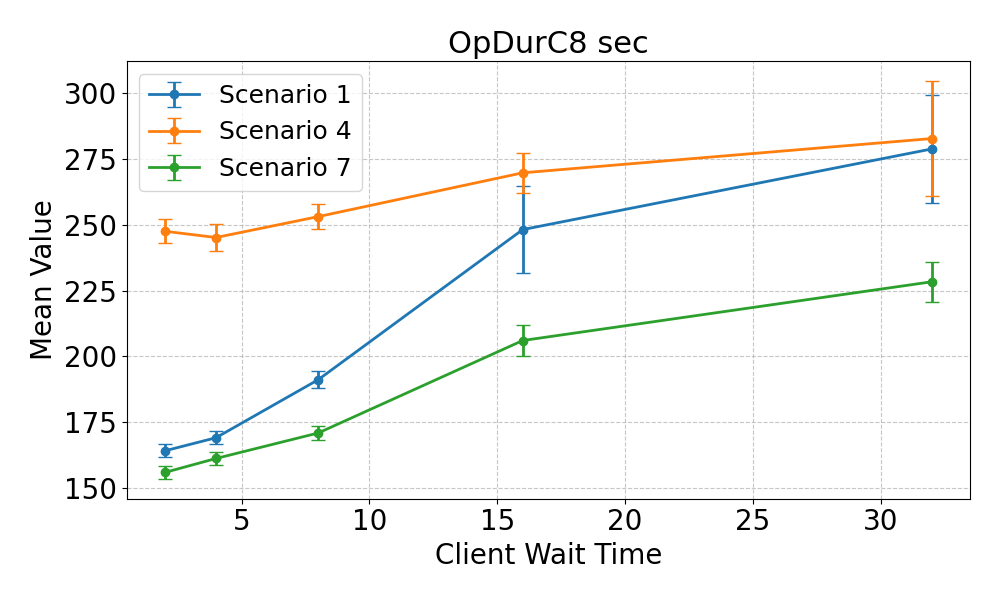}
    \caption{We vary the mean wait time (sec) in Alice's markov model; longer waits between posts means lower performance. We observe the effect on KL divergence bounds (left), on probability of alarm for the cumulative C8 detector (middle), and on operating duration (right). The bar around the datapoints is the radius. In the KL bounds figure on the left, scenario 1 and 4 datapoints overlap. The threshold $N=80$ for the C.8 detector is fixed.}
    \label{fig:cw-vs-threshold}
\end{figure*}
We quantify undetectability--performance tradeoffs across three representative scenarios (Scenarios~1, 4, and~7), which differ in network loss and background traffic intensity (Table~\ref{table:scenarios}). In each scenario, Alice sends a fixed set of covert files while we vary a single application parameter, the mean client wait time between successive HTTP requests in Alice's Markov posting model. Decreasing this wait time increases the rate at which covert traffic is injected and therefore increases goodput.

For each configuration, we audit detectability using a fixed passive detector $\varphi$ and estimate its true positive rate and false positive rate under the HCS and ordinary models, respectively. In this experiment, $\varphi$ is the cumulative C8 detector, which alarms when the number of observed HTTPS requests exceeds a fixed threshold $N$ within the observation horizon (Section~\ref{sec:passive_adversary_description}). We convert the resulting SMC estimates of $\mathrm{TPR}_\varphi$ and $\mathrm{FPR}_\varphi$ into a certified lower bound on $\mathrm{KL}(Q_t \| P_t)$ using the auditing procedure in Section~\ref{subsec:undetectability-auditing}. Figure~\ref{fig:tradeoffs-s7} plots, for each scenario, the certified KL lower bound as a function of the achieved goodput.

Across all three scenarios, increasing goodput generally increases statistical distinguishability between HCS and ordinary executions. Intuitively, more aggressive sending shifts the detector-relevant traffic statistics away from their ordinary distribution, which increases $\mathrm{TPR}_\varphi$ relative to $\mathrm{FPR}_\varphi$ and therefore increases the certified KL lower bound. The scale of the bound varies substantially across scenarios, reflecting the dependence of detectability on network conditions and background traffic.

These results also illustrate how divergence-based undetectability claims can be audited by sweeping a deployment parameter. Suppose a deployment claims to provide $(\mathrm{KL}, 0.01)$-undetectability under a specified scenario configuration (e.g., Scenario~7) at a given operating point. If the certified lower bound in Figure~\ref{fig:tradeoffs-s7} exceeds $0.01$ at that operating point, the claim is falsified under the modeling assumptions and detector specified in Sections~\ref{sec:passive_adversary_description} and~\ref{subsec:undetectability-auditing}. Conversely, a lower bound below $0.01$ does not refute the claim, but identifies operating regimes in which the measured detector behavior is at least consistent with the target.

Finally, we observe that the KL divergence lower bound is very high for the other scenarios for the C8 detector, and we do not show these results. This is expected given that the HCS is not private (and adversary can easily distinguish between ordinary and HCS even with these simple detectors). \\

\noindent\textbf{Detector parameter sweep.} In Figure~\ref{fig:tkl-vs-threshold}, we sweep the moving-average detector threshold multiplier $k$, which corresponds to evaluating a family of statistical tests with different operating characteristics. Each choice of $k$ induces a different binary classifier $\varphi_k$ with its own $\mathrm{TPR}_{\varphi_k}$ and $\mathrm{FPR}_{\varphi_k}$, and therefore its own lower bound on $\mathrm{KL}(Q_t \| P_t)$. The key observation is that different thresholds produce substantially different KL lower bounds. In particular, more conservative thresholds (larger $k$) reduce the false positive rate and can yield significantly larger KL lower bounds, even if the true positive rate also changes. Since each threshold defines a valid detector, we may view this sweep as auditing the HCS against a suite of statistical tests. An audit can then report the maximum lower bound across this family as the strongest divergence evidence obtained from the tested detectors. \\

\noindent\textbf{Application parameter sweep.}
In Figure~\ref{fig:cw-vs-threshold}, we vary Alice’s performance by changing the mean wait time (sec) between successive posts in Alice’s Markov model. Longer waits between posts increase the time required to send a file and therefore reduce goodput. In this experiment, the cumulative C8 detector threshold is fixed. For each wait time, we measure the certified KL divergence lower bound (left), the probability of alarm for the C8 detector (middle), and the operating duration until detection (right).

As Alice becomes more aggressive, corresponding to shorter wait times, the detector achieves greater separation between $\mathrm{TPR}$ and $\mathrm{FPR}$, which produces larger certified KL lower bounds. Conversely, as the wait time increases and traffic more closely resembles ordinary behavior, the KL lower bound decreases. The operating duration increases with wait time, reflecting that less aggressive sending delays or reduces detection events.

The key takeaway is that divergence is determined jointly by application behavior and adversarial test configuration. Even with a fixed detector threshold, performance tuning alone can move the system between operating regimes that yield large certified distinguishability and regimes that yield substantially smaller divergence evidence.

\subsection{Semantic Alignment Validation}\label{subsec:semantic_alignment}
The core of this audit consists in comparing metrics collected from real systems on a testbed against those derived from SMC analysis of our model.

\subsubsection{Quantification}\label{subsec:semantic_alignment:quantification}
\begin{figure}[t!] 
    \centering 
    \begin{subfigure}{0.5\textwidth}
        \centering
	    \includegraphics[width=\linewidth]{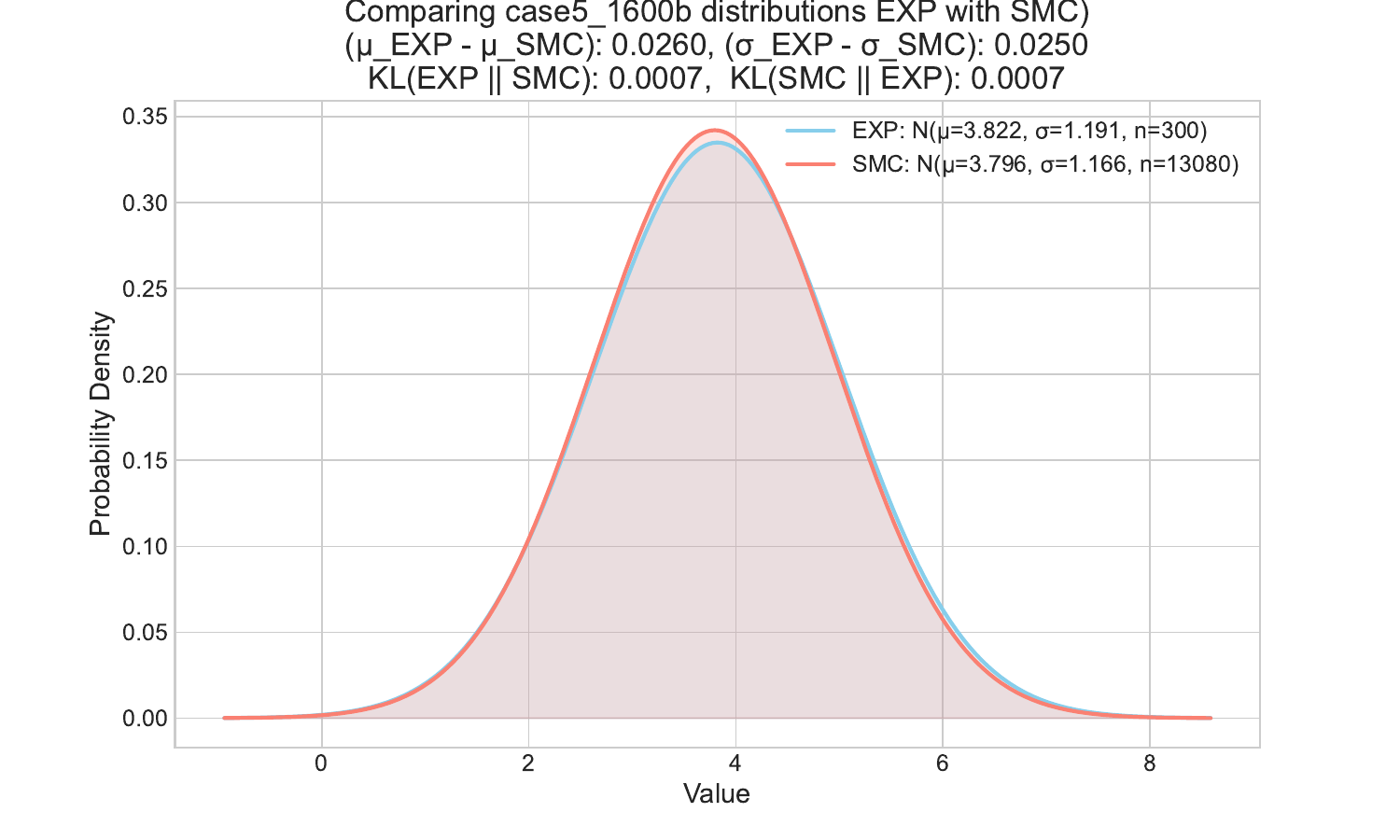}
	    \label{fig:cp1_case5_latency}
    \end{subfigure}
    \hfill 
    \begin{subfigure}{0.45\textwidth}
        \centering
        \includegraphics[width=\linewidth]{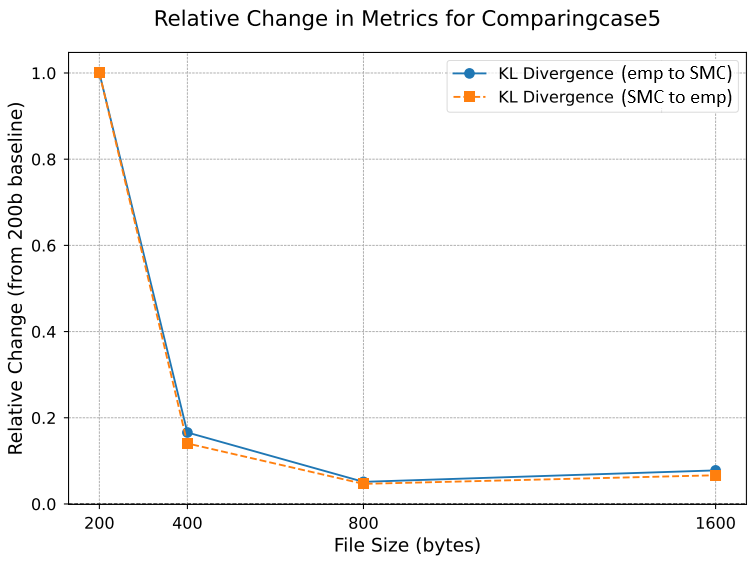}
        \label{fig:cp1_case5_kl}
    \end{subfigure}
    \caption{Close Iodine alignment from the latency distributions between Maude model of Iodine and empirical measurements (top), the normalized KL divergence as the file to transmit increases in size 1,600B (bottom).}
    \label{fig:cp1_case5}
\end{figure}

Our first effort to quantify semantic alignment focuses on the Iodine control channel only: we modeled and derived expected values of latency and goodput metrics for elemental Iodine behavior.
Our goal is to validate Iodine performance guarantees, \emph{i.e.,} to demonstrate empirically that model guarantees transfer to Iodine as actually implemented.
We begin by characterizing the latency to transfer a 1,600B file.
Given maximum query sizes of 100B, the file will require 16 DNS (encoded) queries.
Figure~\ref{fig:cp1_case5} (left) shows close agreement between our model and empirical measurements, with around 26ms difference in the means, otherwise similar distributions, and very low KL divergence (near zero).~\footnote{This KL divergence is empirically computed on two distributions each characterized by a mean and a standard deviation, and is only intended as a proxy for distance.}
With tight alignment from the performance metrics we measure like latency and goodput, we can observe that exfiltration latency is driven by the number of queries required to send the file in the absence of loss on the links, and loss and retransmissions in the presence of loss.
Figure~\ref{fig:cp1_case5} (right) shows the normalized KL divergence relative to file size. Relative KL divergence decreases and remains very low with increasing file size.

We further substantiate semantic alignment of the entire HCS (Iodine and Destini-Mastodon) by comparing metrics extracted from the testbed and Maude's HCS model SMC results across the scenarios of Section~\ref{subsec:scenarios_description}. For each scenario, empirical results are aggregated from 1,024 runs transmitting between five and ten files. SMC results are the expected value of the metrics after 300 Monte Carlo simulations.
\begin{figure*}[t!] 
    \centering 
    \begin{subfigure}{0.45\textwidth}
        \centering
	    \includegraphics[width=\linewidth]{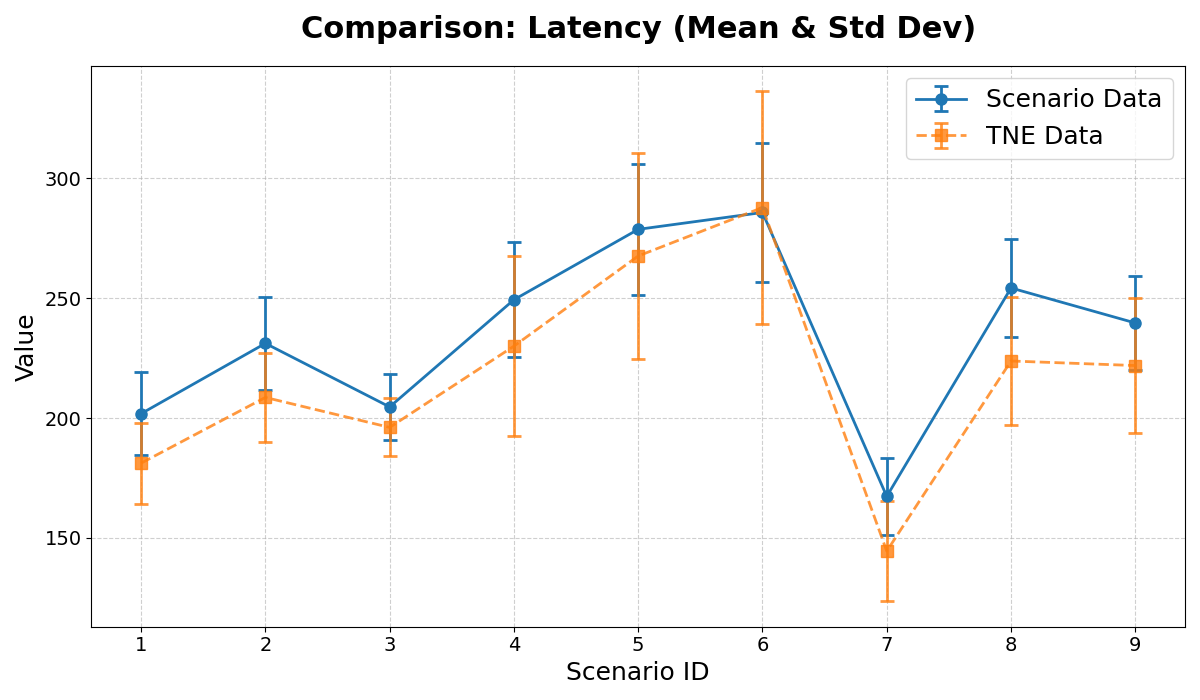}
        \label{fig:latency_scenarios}
    \end{subfigure}
    \hspace{0.cm}
    \begin{subfigure}{0.45\textwidth}
        \centering
        \includegraphics[width=\linewidth]{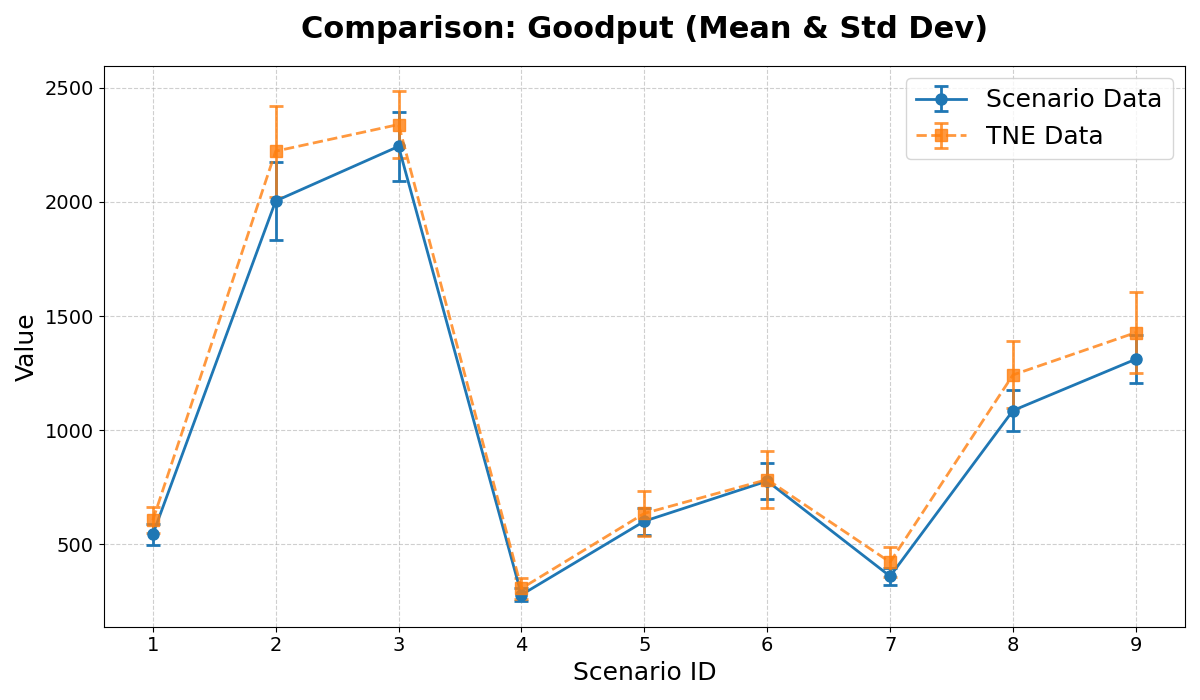}
        \label{fig:goodput_scenarios}
    \end{subfigure}
    \hfill
    \begin{subfigure}{0.45\textwidth}
        \centering
        \includegraphics[width=\linewidth]{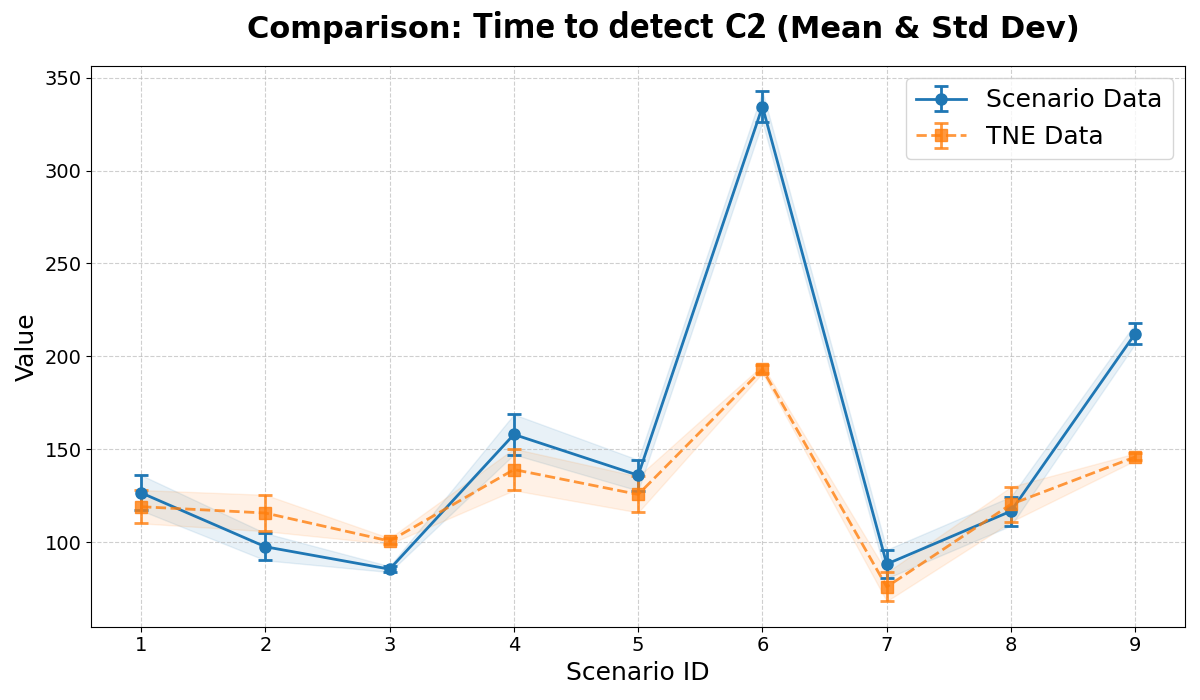}
        \label{fig:timetodetectC2_scenarios}
    \end{subfigure}
    \begin{subfigure}{0.45\textwidth}
        \centering
        \includegraphics[width=\linewidth]{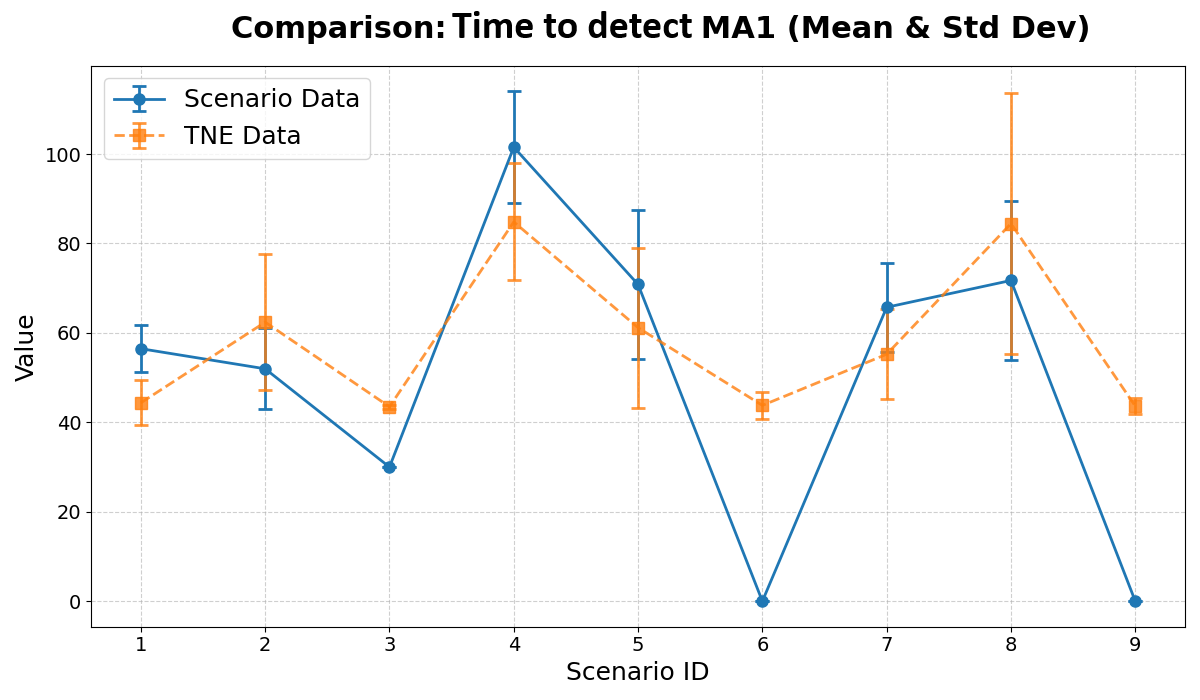}
        \label{fig:timetodetectMA1_scenarios}
    \end{subfigure}
    \caption{Semantic alignment verification using latency (top left), goodput (top right), Op duration (time-to-detect) monitoring DNS queries counts (bottom left), and Op duration monitoring DNS query rate (bottom right) metrics between Test and Evaluation (TNE) and Maude. Error bars show one standard deviation from the mean. Our HCS model comes close (around one standard deviation) to the statistics collected on the testbed.}
    \label{fig:latency_goodput_scenarios}
\end{figure*}
Figure~\ref{fig:latency_goodput_scenarios} shows that the Maude model tracks the testbed metrics well, with our model mostly falling within one standard deviation of the testbed-collected statistics for latency and goodput, and at most 13.6\% apart for latency, and 17.1\% for goodput. For the Operating duration (time-to-detect) metric (Figure~\ref{fig:latency_goodput_scenarios}, bottom left), our model aligns with empirical data well, except for scenarios 6 and 9. However, these do not necessarily invalidate our model per-se. Recall that scenarios 3, 6, and 9 only differ from their group's other scenarios in how many background traffic generators they instantiate, which suggests it is our model for background traffic in the presence of loss that is inaccurate, not our HCS model. Figure~\ref{fig:latency_goodput_scenarios} bottom right shows the same alignment for all scenarios but 6 and 9.

\section{Discussion and Conclusions}\label{sec:conclusion}


In this paper we introduced the \MH\ executable modeling
and analysis framework and described its use for analyzing
performance and indistinguishability properties of hidden
communication systems. We propose a method to quantify
indistiguishabiliy of hidden communications from the underlying
communications. We use this to develop a formal method to audit
measures of the differences between the observations of the hidden
communications for a  protocol $W[\mathcal{R}]$
and the analogous observations for the underlying
protocol $\mathcal{R}$ which is used
by $W[\mathcal{R}]$ to hide its communications.

We evaluate the \MH\ framework, auditing and
tradeoff methods on a collection  scenarios
representing real-world hidden communication systems,
including DNS tunneling and posting pictures with
embedded information on a whiteboard such as Mastodon.

The models presented here were developed in a project
where the aim was to demonstrate that
formal models of existing  hidden communication system implementations
can provide realistic
performance and detectability analyses, thus reducing
the need for lengthy testbed simulations.
Beyond the basic challenge of modeling such complex systems,
this meant capturing the behavior of existing implementations
that are full of quirks, unspecified features, and bugs.
We were able to show that formal analysis based on 
the Maude-HCS models replicated with substantial faithfulness
experimental results in many cases; but it is not yet clear  in general 
what it means for a formal model to align with such implementations.   

An important next step is to \emph{invert the process}:  to first (i) design
and analyze a new hidden communication system using the \MH\ framework,
to formally analyze its  performance and undetectability properties,
and to improve and  fine-tune the system's design to meet 
desired performance and indistinguishability goals
(which is orders of magnitude easier and more informative
than evaluating, optimizing and fine-tuning an actual implementation);
and then (ii) generate an implementation of the fine-tuned and
thoroughly analyzed HCS protocol design
from its formal  model.  For greater assurance,
by using code generation methods similar to the
one proposed in \cite{D-transf-NASA}, it may even be possible to
generate a high-quality  implementation in an  automatic and
correct-by-construction way.

We developed the current models by carrying out by hand the
theory transformations needed to specify and analyze hidden
communication protocols, and 
to convert non-deterministic models into
purely probabilistic models that
can be used for SMC analysis by Monte Carlo simulation.
An important next step is to implement the already existing
algorithmic definitions of the different formal
patterns in \cite{timedPTran}
used in the \MH\ modeling and analysis
tool chain: automating 
 these transformations will make it much easier for communication protocol
 designers to use \MH\ in order to
 reach high-quality designs based on solid performance
 and security estimations of their behavior \emph{before} their
 designs are implemented.

Simulating and analyzing performance involves considering
scenarios with large numbers of actors, some engaged
in normal communication, some carrying out hidden communication.
For formal models and their SMC analysis
to scale up to large scenarios, together with SMC parallelization
methods such as those in \cite{alturki_pvesta_2011}, which
are already supported by QMaude,
we also need to develop
\emph{abstraction methods} that reduce the state space while
still capturing all the sytem's relevant features.  
\section*{Acknowledgments}
DISTRIBUTION STATEMENT A: Approved for public release; distribution is unlimited. This material is based upon work supported by the Defense Advanced Research Projects Agency (DARPA) under Contract No HR001125CE019.

\bibliographystyle{ACM-Reference-Format}
\bibliography{refs, zotero}

\appendix

\section{Modeling Hidden Communication in Maude-HCS}\label{sec:rtt-section}

We illustrate the basic ideas for modeling and analyzing
hidden communication in Maude-HCS with a simple round
trip time protocol, RTT, and a transformed version,
WRTT, that uses RTT to hide the transmission of a stream
of bytes, for example a stream of sensor measurments. An
RTT session has two actors A and B. Actor A periodically
sends an RTT request to B containing a timestamp
\texttt{ts(T)}, a digital representation of the current
time. B responds to the request with a pair consisting
of the timestamp received from A and the timestamp
corresponding to the current reading of the clock by B.
When A receives the response, the roundtrip time (the
current clock reading minus the request timestamp) is
computed and added to the senders local list of round
trip times. RTT is transformed to transmit hidden
information from the B to the A. Using a formal pattern
the actors A and B are wrapped in meta actors that do the
embedding and extracing of hidden information, leaving A and B
unchanged. We present the result of a further
transformation (optimization) that merges the attributes
of the base and meta actors and combines the effects of
the base and meta actor rules. Thus, in the WRTT protocol B is modified to
embed a byte of hidden information in the lower 8 bits of
response timestamp, and A is modified to extract that hidden
byte in addition to computing the round trip time, which,
as we shall see, is indeed the \emph{correct} round trip time
in spite of B's modified timestamp.

The conjecture is that the hidden communication with WRTT is
difficult to detect since it doesn't change the message
size or timing of RTT, and the lower order bits of actual
timestamps are likely to be unpredictable due to clock
reading randomness. We can study this conjecture using
Maude-HCS, by specifying the RTT protocol and its
transformation to WRTT, and further specifying what can
be observed and how the adversary might use what is
observed.

The Maude-HCS model of the RTT protocol
is a PMaude model of the form $\mathcal{R}{_{\mathit{rtt}_{\Pi}}}$,
which is the result of applying the timed $P$ transformation
from \cite{timedPTran} to the RTT protocol $\mathcal{R}_{\mathit{rtt}}$
by choosing probability distributions $\Pi$ for message delays.
RTT itself is modeled as a real-time rewrite theory  $\mathcal{R}_{\mathit{rtt}}$ that
provides RTT's Real-Time Maude  (RTM) semantics\cite{journ-rtm} and includes 
definitions of the attributes of the two actors, the 
rules defining the actors' behavior.  $\mathcal{R}_{\mathit{rtt}}$ is then
modularly extended with a
specification
of adversary capabilities, and and specification of
analysis scenarios.  

The specification of $\mathcal{R}_{\mathit{rtt}}$
uses an abstract
notion of time \texttt{(sort Time)} which can be
instantiated to natural numbers, non-negative rationals,
or floating point numbers for different analyses.
Floating point approximation to real numbers is used for statistical analysis.

In the following we explain the Maude rules for $\mathcal{R}_{\mathit{rtt}}$
and show how  $\mathcal{R}_{\mathit{rtt}}$ is transformed 
into the WRTT protocol $\mathcal{R}_{\mathit{wrtt}}$.
We then describe how observables are collected
to support analysis by statistica model checking and 
auditing.\footnote{The specification and analysis scenarios are available in the Maude-HCS repository.}

The periodic behavior of A is specified using Real Time
Maude (RTM) timers. A timer is represented by a term of
the form \texttt{[T,P,data]} where \texttt{T} has sort
Time (the time remaining), \texttt{P} has sort Nat (the
period), and \texttt{data} is auxiliary data, not used
in the RTT protocol.

There are two rules for the behavior of the RTT sender A
and one rule for the receiver B.

The sender rule \texttt{[sndRttReq]} specifies sending
an RTT request \texttt{rttReq(ts)} when the timer
reaches zero.

\begin{small} 
\begin{verbatim} 
crl [sndRttReq] : 
  < A : Snd | timers: [tt ; per ; data], tsq: tsl,
              clock: T, stopT: Tstop, rcv: B, atts > 
   => 
  < A : Snd | timers: [tt0 ; per ; data], tsq: (tsl ; ts), 
    clock: T, stopT: Tstop, rcv: B, atts > 
 [ (to B from A : rttReq(ts)) ] 
 if tt == zero 
 /\ tt0 := (if T ge Tstop then infty else nat2t(per) fi) 
 /\ ts := t2ts(T) . 
\end{verbatim} 
\end{small}

\noindent
The term \texttt{t2ts(T)} represents the
timestamp obtained by reading a clock at time \texttt{T}.
The sender A updates its timestamp queue attribute \texttt{tsq:}  by
appending \texttt{ts} to the list of previously sent timestamps
\texttt{tsl}.  This allows A to connect each response from B with
a specific request from A.
A also resets its
timer either to \texttt{infty} (the infinite time)
if the clock value \texttt{T} is greater than the stop 
time, \texttt{Tstop}, or to the request period, \texttt{per}, (converted from a natural number to a time, nat2t(per), otherwise.)
Setting the timer value to \texttt{infty} is
equivalent to turning the timer off, as decrementing \texttt{infty}
by a finite value remains \texttt{infty}.  In this case
the sender will stop sending.

The term \texttt{[(to B from A : rttReq(ts))]} is an RTM message scheduling task.  Such tasks are rewritten to delayed messages \texttt{dly(M,dT)} by RTM rules.  Delayed messages
remain in the configuration until the amount of time \texttt{dT} 
has elapsed. 

The sender rule \texttt{[rcvRttResp]} specifies sender A's behavior when a
response \texttt{rttResp(ts,ts0))} arrives from the receiver B.
The computed round trip time, \texttt{computeRtt(T,ts))}, is
appended to the attribute \texttt{rttq:}.  A then
waits for the timer to reach zero to send the next rtt request.
The passage of time is specified by the RTM \texttt{tick}
rule that computes the minimal time that must elapse before some
action (i.e., some rule firing) is enabled.

\begin{small}
\begin{verbatim}
rl[rcvRttResp]:
 < A : Snd | clock: T, rcv: B,  timers: [tt0 ; per ; data],
             tsq: (tsl0 ; ts ; tsl1), rttq: rttl  , atts > 
(to A from B : rttResp(ts,ts0))
=>
< A : Snd | clock: T, rcv: B, timers: [tt0 ; per ; data],
            tsq: (tsl0 ; tsl1),
            rttq: (rttl  ; computeRtt(T,ts)),  atts >  .        
\end{verbatim} 
\end{small}

The rule \texttt{[rcvRttReq]} specifies that when a request
\texttt{rttReq(ts))} from A is received by the receiver B,
then B sends a response message of the form \texttt{rttResp(ts,t2ts(T))}
 to A containing the received timestamp and
its current timestamp.
B does not change its state.

\begin{small}
\begin{verbatim}
rl[rcvRttReq]:
 < B : Rcv | clock: T, snd: A, atts >
  (to B from A : rttReq(ts))
=>  
 < B : Rcv | clock: T, snd: A, atts >
[ (to A from B : rttResp(ts,t2ts(T))) ] .
\end{verbatim} 
\end{small}

Guided by the formal pattern for WRTT, rules from the RTT protocol $\mathcal{R}_{\mathit{rtt}}$
are transformed into those of $\mathcal{R}_{\mathit{wrtt}}$
to carry
hidden information as follows.  The sender A and the receiver B
share a common attribute \texttt{ctr: j}, with \texttt{j} initially 0,
that they increment each round.  This makes possible for A and B to generate
a shared sequence of (pseudo) random numbers.

The sender rule \texttt{[sndWRttReq]} is unchanged.
The receiver rule is transformed: B now generates a random byte, denoted by
\begin{align*}
    \texttt{b(sampleUniWithIntX(i,256))}
\end{align*}
and embeds it
into its timestamp using a function \texttt{embed(ts,j,byte)}.
The new rule also increments the counter.

\begin{small}
\begin{verbatim}
crl[rcvWRttReq]:
 < B : bWRcv | clock: T, snd: A,  bctr: i,   ctr: j, 
              byteList: bytel, atts >
  (to B from A : rttReq(ts0))
=>  
 < B : bWRcv | clock: T, snd: A, bctr: (s i), ctr: (s j), 
             byteList: (bytel ; byte), atts > 
   [(to A from B : rttResp(ts0,ts1))]
  if ts := t2ts(T)
  /\ byte := b(sampleUniWithIntX(i,256))
  /\ ts1 := embed(ts,j,byte) .
\end{verbatim} 
\end{small}

The sender rule to receive the RTT response is modified
to extract the hidden information, add the result byte
to its \texttt{byteList:} attribute, and increment
its counter.

\begin{small}
\begin{verbatim}
crl[rcvWRttResp]:
    < A : bWSnd | clock: T, rcv: B, rttq: rttl,  ctr: j, 
           tsq: (tsl ; ts0 ; tsl1), byteList: bytel, atts >   
   (to A from B : rttResp(ts0,ts1))
   =>
    < A : bWSnd | clock: T, rcv: B,
                  rttq: (rttl ; computeRtt(T,ts0)), ctr: s j,
                  tsq: (tsl ; ts0 ; tsl1),
                  byteList: (bytel ; byte),  atts >     
 if byte := extract(ts1,j) .
\end{verbatim} 
\end{small}

The embedding function replaces the low order 8 bits of
the timestamp (viewed in binary representation) by the
result of xoring the byte with the low order 8 bits of the
random number given by the counter value, \texttt{j}. The
extracting function xors the low order 8 bits of the
timestamp with the same random number (using the value of
the syncronized counter). Since this the same random
number used by the receiver, the embed byte is
recovered.

Once we have specified the real-time rewrite theories
$\mathcal{R}_{\mathit{rtt}}$ and $\mathcal{R}_{\mathit{wrtt}}$
we can use the timed $P$ transformation in
\cite{timedPTran} (choosing the same $\Pi$ for both protocols), as well as the $\mathit{Sim}$ and $M$
transformations described in Section \ref{sec:background}, to perform
SMC analysis of RTT and WRTT by means of the executable PMaude theories $M(\mathit{Sim}(\mathcal{R}_{\mathit{rtt}}))$ and $M(\mathit{Sim}(\mathcal{R}_{\mathit{wrtt}})$,
or, as explained below, of their extensions when an \emph{observer} is added to the state.

Configurations consisting of a sender A and receiver B can be
used to generate sample execution traces that can be
inspected or subject statistical model checking, for
example to determine the expected roundtrip time over a
given time interval.

To analyze properties based on observables and measure
differences between the RTT and WRTT scenarios, we add an
observer actor to the configuration. The observer
attributes are chosen to model information an
observer/adversary can obtain under different conditions.
To study the effect of the hidden communication, we define
an observer that collects the list of responses that the
sender actor receives. Such an observer has the form

\begin{small}
\begin{verbatim}
    < obs : Observer | sent: timeMsgList >
\end{verbatim} 
\end{small}

The observation mechanism is specificed as a set of
functions executed by the RTM rules. The RTM rule
\texttt{[sendDelayed]} for converting scheduling tasks
\texttt{[M]} to delayed messages \texttt{dly(M,T)} also
executes the function \texttt{sendTimeAction(conf0, M,T)}
where \texttt{conf0} is the rest of the configuration. The
rule \texttt{[msgArrival]} that rewrites a message with
zero delay, \texttt{dly(M,zero)}, to a receivable message
\texttt{M} also executes the function
\texttt{arrivalTimeAction(conf,M}). The RTM rule 
\texttt{[infty]} applies when the minimal time 
to advance is \texttt{infty}, indicating the end of the execution.
This rule applies the \texttt{finalAction} to the
configuration and terminates the execution.
The \texttt{finalAction} function may collect
additional information from the actors and may add summary
information to the observer. By default, all
these functions are the identity on configuration. Each
system analysis should define them as needed.
 
In the RTT/WRTT examples, \texttt{arrivalTimeAction} is
defined to add the pair \texttt{(T,M)} to the observer
\texttt{sent}: attribute, thus collecting all the responses
received with the time of receipt. Thus, when the stop time
is reached, the observer object will have the form:

\begin{small}
\begin{verbatim}
    < obs : Observer | sent: (T1,M2) ; ... ; (Tk,Mk) >
\end{verbatim} 
\end{small}

\noindent
where \texttt{M1 ... Mk} are the  responses received by
the sender.  

In the RTT/WRTT cases, the
\texttt{finalAction} function adds three new attributes to
the observer. The value of the \texttt{enList} attribute
is a list of the entropies of the 8 low order bits receiver timestamps (the second timestamp in the
response pair) in the \texttt{sent:} attribute.  The value of the \texttt{enAv}: attribute is the
average of the values in the entropy list. The value of
the \texttt{rttAv:} is the average of the round trip times
recorded by the sender in the \texttt{rttq:} attribute.

For example, the results below are from eight successive runs
of the PMaude models of RTT and WRTT with parameters: startTime
999., stopTime 10999., Period 500, and link delay between
sender and receiver sampled from a normal distribution
with mean 50.0 and standard deviation 10.0.

\begin{small}
\begin{verbatim}
RTT
enAv: 6.0863390095915237e-1, rttAv: 9.7904761904761898e+1
enAv: 7.06767431818905e-1,   rttAv: 9.8571428571428569e+1
enAv: 6.5459027078376264e-1, rttAv: 9.8904761904761898e+1
enAv: 5.9602399389404914e-1, rttAv: 9.8619047619047606e+1
enAv: 7.0761801320459572e-1, rttAv: 9.7809523809523796e+1
enAv: 6.2943397982393068e-1, rttAv: 9.8049999999999997e+1
enAv: 7.8639975258512451e-1, rttAv: 9.7299999999999997e+1
enAv: 5.463652286124957e-1,  rttAv: 9.7949999999999989e+1
\end{verbatim} 
\end{small}

\begin{small}
\begin{verbatim}
WRTT
enAv: 7.2691476093154683e-1, rttAv: 9.7904761904761898e+1
enAv: 7.4344054910605573e-1, rttAv: 9.8571428571428569e+1
enAv: 7.2093973227398722e-1, rttAv: 9.8904761904761898e+1
enAv: 6.981128141131786e-1,  rttAv: 9.8619047619047606e+1
enAv: 7.2158497284097312e-1, rttAv: 9.7809523809523796e+1
enAv: 7.1022824513931959e-1, rttAv: 9.8049999999999997e+1
enAv: 7.1032439283023907e-1, rttAv: 9.7299999999999997e+1
enAv: 7.0959202890719641e-1, rttAv: 9.7949999999999989e+1
\end{verbatim} 
\end{small}

Informally it appears that the round trip time averages
have similar distributions in the two cases, while the
entropy seems to be slightly higher in the WRTT
case; but a thorough application of SMC is still needed to confirm
this impression.

For detectability we can compute the expected value
of the entropy at some high confidence for each case.
We could also pick a threshold, for example .7, and 
carry out true/false positive analysis.

\section{Expressing Mixed Properties}\label{sec:quatex-example}
It is intuitive to express the operating duration quantitative property as a mixed properties in QuaTEx.
To specify the operating duration (time of detection) for the C2 detector,

\begin{small} 
\begin{verbatim} 
OpDurationC2() = 
 if 
    (s.rval("ToDNQueryPostNAT(C,NTHRESHOLD)") == 0.0) 
 then 
    discard
 else 
    (s.rval("ToDNQueryPostNAT(C,NTHRESHOLD) - 
                    ExfilStartTime(getMonitor(C))"))
 fi;
 
 eval E[OpDurationC2()];
\end{verbatim} 
\end{small}

The \verb|ToDNQueryPostNAT(C,NTHRESHOLD)| computes the total number of post-NAT DNS queries in the adversary's observation (which is included in the configuration \verb|C|), scanning the trace from the beginning, and at the point in time $T$ when the threshold \verb|NTHRESHOLD| is exceeded it returns the difference between $T$ and the time when Alice started exfiltrating its files (computed by \verb|ExfilStartTime| over the monitor object which is also in the configuration).

This is a {\em conditional expectation}, which focuses the measure on sample runs where the alarm actually fires.
Sample runs where the threshold is not exceeded are discarded.
QMaude reports the expected value with its confidence radius, and the number of simulation runs that were discarded. 
This property is precise, and the measure is statistically grounded.

\end{document}